\documentclass[10pt,journal]{IEEEtran}

\usepackage[T1]{fontenc}
\usepackage[utf8]{inputenc}
\usepackage{booktabs}
\usepackage{array}
\usepackage{tabularx}
\usepackage{multirow}
\usepackage{graphicx}
\usepackage{xcolor}
\definecolor{bindingblue}{HTML}{3F72B5}
\definecolor{rulegray}{HTML}{66717F}
\usepackage{tikz}
\usepackage{amsmath}
\usepackage{url}
\usepackage{algorithm}
\usepackage{algpseudocode}
\usepackage[hidelinks]{hyperref}
\usepackage{eso-pic}
\hypersetup{pdftitle={Beyond Approved Actions: Runtime Validation of Persistent Outcomes in Agent Workflows},pdfauthor={Haoran Zhang, Hengtong Zhang, Zhiyu Liang, Yu Yan, Decheng Zuo, Hongzhi Wang}}

\newcommand{\sys}{EffectMatch}
\newcommand{\term}[1]{\texttt{#1}}
\newcommand{\gaplabel}[1]{\textbf{#1)}}
\newcommand{\modulebadge}[1]{\tikz[baseline=(badge.base)]{\node[circle,fill=black,text=white,font=\scriptsize\bfseries,inner sep=0pt,minimum size=1.15em] (badge) {#1};}}

\begin{document}
\AddToShipoutPictureFG*{\AtPageLowerLeft{\put(52,20){%
\parbox[b]{508pt}{\normalfont\fontsize{7}{8}\selectfont This work has been submitted to the IEEE for possible publication. Copyright may be transferred without notice, after which this version may no longer be accessible.}}}}

\title{Beyond Approved Actions: Runtime Validation of Persistent Outcomes in Agent Workflows}

\author{Haoran~Zhang, Hengtong~Zhang, Zhiyu~Liang, Yu~Yan, Decheng~Zuo,
        and~Hongzhi~Wang,~\IEEEmembership{Senior Member, IEEE}%
\thanks{Haoran~Zhang, Hengtong~Zhang, Zhiyu~Liang, Yu~Yan, Decheng~Zuo, and Hongzhi~Wang are with the Faculty of Computing, Harbin Institute of Technology, Harbin 150001, China.
E-mail: zhr@stu.hit.edu.cn, \{hengtong, zyliang, yuyan, zuodc, wangzh\}@hit.edu.cn.}%
\thanks{Corresponding author: Hongzhi~Wang}}

\markboth{Preprint -- Submitted to IEEE Transactions on Software Engineering}%
{Beyond Approved Actions: Runtime Validation of Persistent Outcomes in Agent Workflows}

\maketitle

\begin{abstract}
Large language model agents increasingly act on software systems, no longer merely generating text but also changing databases and online services. However, an approved database update may succeed yet leave an unapproved notification because execution can produce persistent effects beyond the requested change. Current safeguards can approve an action or record its aftermath, but without checking the persistent result before continuation, an unapproved outcome can be accepted as success and propagated to later steps. We present \sys{}, a runtime that collects persistent changes within a controlled execution boundary and compares them with what the application approved for the current state and execution. The comparison governs commit and dependent execution. In comparative evaluation on 206 public business tasks, \sys{} preserved all clean executions and prevented all tested incorrect commits. Six 20-run ablations exposed the failure caused by each removed mechanism, while 80 task-topology cases preserved truthful handoffs and blocked invalid continuation. Together, these results show that \sys{} blocks the silent acceptance and downstream propagation of persistent outcomes inconsistent with application approval.
\end{abstract}


\IEEEpeerreviewmaketitle

\section{Introduction}
\label{sec:intro}

\IEEEPARstart{L}{arge} language model agents are moving from content-generation tools into active participants in software workflows. They can read business records and then modify databases, send messages, or invoke external services; whether these operations can be delegated safely will determine how far agents can enter real software systems. In the intuitive execution model, an application reviews a call proposed by an agent, the runtime checks its target, arguments, and authority, and the workflow continues when the backend reports success. This model works naturally when a call produces only the result it directly describes. Once agents modify persistent state, however, that assumption no longer holds: a call can be approved and return success while leaving an outcome the application never approved, and later tasks may proceed without knowing it.

Existing research and deployed systems already protect agent actions before, during, and after execution. They check whether requests comply with authority and policy~\cite{south2025delegation,qin2026airguard,ma2026secureclaw}, prevent operations from completing only partially or being repeated by retries~\cite{gray1992transaction,helland2012idempotence,zhang2020beldi}, and later reread or record system state~\cite{santosgrueiro2026temporary,xu2026caplease,cohen2026witness}. These safeguards are necessary: they can establish that a call was permitted, that execution did not obviously fail, or that a record remains for later inspection. They still do not establish that every change ultimately left in the system was within the application's approval. A call can therefore pass its checks and return success while also producing an unapproved change.

\begin{figure}[t]
\centering
\includegraphics[width=\linewidth]{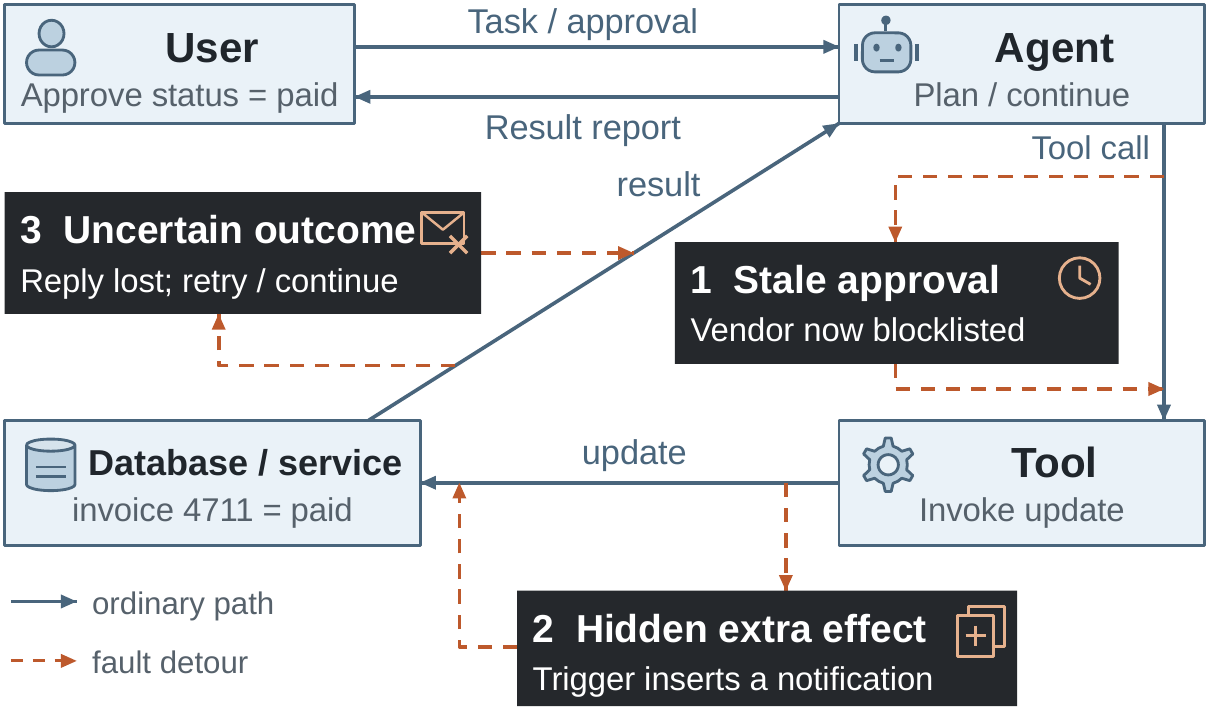}
\caption{Normal and fault-detour paths can reach the same downstream component, exposing stale approval, hidden effects, and uncertain completion.}
\label{fig:gaps}
\end{figure}

Figure~\ref{fig:gaps} follows an agent processing invoice 4711. The application approves changing its database status to \emph{paid}; on the ordinary path, the backend performs the update, reports success, and the workflow moves to dependent reconciliation tasks. Yet approval, execution, and continuation can still come apart in three ways:
\begin{enumerate}
\renewcommand{\labelenumi}{\gaplabel{\arabic{enumi}}}
\setlength{\itemsep}{0.15em}
\setlength{\parsep}{0pt}
\setlength{\topsep}{0.25em}
\item The approval may become stale before execution. If the vendor is blocklisted after approval, the same update no longer has current authority.
\item A database trigger can insert an unapproved notification together with the requested update, so the call returns success but leaves an additional persistent result.
\item A remote payment request may be dispatched while its response is lost. Retrying it or releasing dependent tasks without knowing the outcome can duplicate the payment and propagate incorrect state.
\end{enumerate}
At the call level, all three executions may appear approved, successful, or dispatched. Safe continuation therefore requires one judgment spanning whether the approval still applies to this execution, what persistent outcome it actually left behind, and whether later work may proceed.

To connect these judgments, the application's approval must continue to constrain both the operation's actual outcome and the handling of that outcome. \sys{} carries the application-approved result through execution. Before execution, the application states the changes that the operation may leave in the system. Specifically, \gaplabel{1} the runtime binds that approval to its source, current state, plan revision, and one occurrence to prevent stale reuse; \gaplabel{2} it first checks whether routes to persistent changes are covered by observation or blocked, then stages or controls the operation, collects the actual changes, and compares them with the approval to expose hidden effects; and \gaplabel{3} it persists the resulting decision to govern commit, retry, and dependent work. Only an execution with a matching outcome and current one-use approval propagates as success. We call this cross-stage constraint over approval, execution, validation, and continuation the \emph{Effect Commit Contract}.

We evaluate \sys{} on public database tasks from MCPMark~\cite{wu2025mcpmark} and business tasks spanning shopping, customer support, and travel. On twelve business tasks with approved outcomes independently specified from requests and initial state before execution, \sys{} preserved all clean executions and rejected the mismatches in all 39 task--fault combinations, each repeated three times. Further experiments cover reference-based comparisons on 206 public tasks, live execution with three agent configurations, cross-process recovery, and multiple database and service backends.

This paper makes three contributions:

\begin{enumerate}
\item We formulate persistent-outcome validation for agent workflows and introduce the Effect Commit Contract, which binds application approval, actual outcomes, and continuation to the same execution.
\item We design and implement \sys{} to establish what an operation actually leaves behind and use the comparison result to govern commit and continuation.
\item We validate outcome checking and execution binding through controlled comparisons and ablations on public tasks, and assess system behavior through live-agent integration, cross-process recovery, and cross-backend experiments.
\end{enumerate}

\section{Execution Model and Persistent-Outcome Problem}
\label{sec:problem}

\subsection{Execution Path and Local Guarantees}

Tool-using agents reach business state through a software execution path. Consider the ordinary handling of invoice 4711. The task owner requests that its database status change from \emph{pending} to \emph{paid}; the agent renders this task as a structured operation; the application or agent runtime decides whether to dispatch it; and a tool connector translates it into a database command, API request, or service invocation. The backend then executes the request, and its response may release reconciliation work, trigger a retry, or enter recovery~\cite{debenedetti2024agentdojo,wu2025mcpmark,zhang2020beldi}. Once committed, the resulting state can outlive both the model turn and the process that initiated it.

We distinguish four logical responsibilities along this path. The task owner or application maintains the business task and its continuation policy. The agent proposes an operation that implements the task. The tool side turns that proposal into a concrete backend request. The backend executes the request, persists its result, and reports status to the workflow. An implementation may co-locate these responsibilities or distribute them across services. This abstraction exposes three interfaces: the business task becomes a software call, the call produces a persistent outcome, and that outcome determines whether later work is released, retried, or stopped.

Each role may satisfy its local interface: the agent emits a well-formed operation, the tool side dispatches it, the backend returns success, and the workflow releases dependent work. These signals describe local progress, but they do not enumerate the persistent outcome. In the running example, a database trigger may insert an additional notification while every interface completes normally. The execution path therefore needs a cross-stage judgment connecting the business outcome that was approved, the persistent changes that actually occurred, and the decision that controls subsequent work.

\subsection{Three Requirements for Persistent Outcomes}

A workflow can rely on a backend result as success only when three requirements hold for the same execution.

\noindent\gaplabel{1}\ The approval must remain valid for this execution. Authorization and runtime enforcement establish who may invoke an operation and which targets or parameters are permitted~\cite{south2025delegation,schneider2000enforceable,ligatti2005edit}. Suppose the vendor is blocklisted after the invoice update was approved. The earlier approval no longer represents the current task state even though the caller and call shape are unchanged. Associating an approval with its designated source, task-state revision, plan revision, and one execution occurrence exposes stale, substituted, and repeated uses.

\noindent\gaplabel{2}\ The complete persistent outcome must be established. Transactions keep a group of backend changes atomic, while readback and auditing expose resulting state~\cite{gray1992transaction}. An interface request, however, names only the operation sent directly to the backend. Database triggers, cascading updates, and server-side logic may create additional persistent changes; requested changes may also be absent or carry different values~\cite{postgresql17triggers}. Reading back the invoice row can therefore confirm the requested update while missing the trigger-written notification. Validation needs a declared observation boundary that covers the relevant state objects and result-propagation channels and provides the evidence required to attribute every change within that boundary to this execution.

\noindent\gaplabel{3}\ The resulting decision must survive failure. A remote payment may be accepted while its response is lost, its initiating process crashes, or its scheduler restarts. Without a durable decision for that occurrence, retry can duplicate the payment and premature continuation can propagate incorrect state. Idempotent requests and recovery records reduce duplicate execution~\cite{helland2012idempotence,zhang2020beldi}; the workflow still needs a persistent result before it retries the operation or releases dependent work.

Current approval determines whether this execution may proceed, the complete outcome records what it produced, and the durable decision controls what happens next. All three refer to the same execution occurrence. The resulting model therefore contains an approved outcome, an established execution outcome, and a persistent classification relating the two.

\subsection{Approved and Established Outcomes}
\label{sec:objects}

\noindent\textbf{Definition 1 (Approved outcome).}
Given task state $s$, candidate execution $a$, and a registered effect profile $p$, the approved outcome $A$ is the finite multiset of persistent changes that the approving party accepts for this execution, expressed at the profile's comparison granularity. Each item records an operation and target at that granularity, together with the result constraints specified by the approval. A relation-write profile compares operations, relations, and write counts; a value-sensitive profile additionally identifies individual resources and resulting values or transitions. The profile fixes the unit of change in advance: write-event records retain each write, whereas state differences capture net changes between the pre-execution and resulting states. Approved and observed outcomes use the same representation and preserve multiplicities within it.

For example, suppose the application approves changing invoice 4711 from \emph{pending} to \emph{paid}. Its approved outcome is
\[
A=\{\!\{\texttt{invoice[4711].status}:\texttt{pending}\!\rightarrow\!\texttt{paid}\}\!\}.
\]
The approving party fixes $A$ before execution from the task requirements, pre-execution state, and business rules. In the example, approval specifies the transition of invoice 4711 from \emph{pending} to \emph{paid} and requires all other business data included in the comparison to remain unchanged. An approval is current for $a$ when it comes from the designated source, refers to the same task-state revision, and its assigned occurrence has not already been used.

\noindent\textbf{Definition 2 (Execution outcome and completeness).}
A registered backend profile $p$ declares a candidate observation boundary $\partial_p$, covering the state objects and result-propagation channels for which it can establish changes. To prevent unapproved changes before durability, the runtime must also certify this boundary for the current execution identity: every directly reachable persistence or dispatch outlet is either mediated by the profile or denied, and an unclassified outlet prevents certification. The execution outcome $E$ is constructed from backend observation evidence by encoding changes attributed to $a$ within the certified boundary in the units specified by $p$, using the same canonical multiset representation as $A$. We write $\mathrm{Complete}_p(E)$ when this representation covers all changes produced by $a$ within the boundary at that granularity, preserves the required comparison fields and multiplicities, and attributes every represented item to $a$.

Continuing the example, suppose a database trigger also inserts a notification:
\[
E=A\uplus\{\!\{\texttt{insert notification(4711)}\}\!\}.
\]
The certified boundary and execution evidence jointly establish completeness; the number of observed records carries no such guarantee. Missing evidence, an unclassified outlet, or boundary-relevant drift leaves $\mathrm{Complete}_p(E)$ unestablished.

\subsection{Outcome Classification and Problem Statement}
\label{sec:capclasses}

\noindent\textbf{Definition 3 (Outcome classification).}
Given $A$, $E$, and profile $p$, $\mathrm{Outcome}_p(A,E)$ takes one of three values:
\begin{equation}
\begin{array}{@{}ll@{}}
\term{EXACT}: & \mathrm{Complete}_p(E)\land E=A,\\
\term{DIVERGED}: & \mathrm{Complete}_p(E)\land E\neq A,\\
\term{INDETERMINATE}: & \mathrm{Complete}_p(E)\text{ unestablished}.
\end{array}
\label{eq:outcome-classification}
\end{equation}
We abbreviate the first condition as $\mathrm{EffectExact}_p(A,E)$. Equality compares all fields and multiplicities selected by $p$; value equality is established by profiles that retain values or their digests. The running example is \term{DIVERGED}: the invoice transition is present, together with an unapproved notification. It would be \term{EXACT} if its complete outcome equaled $A$, and \term{INDETERMINATE} if the registered profile could not establish whether it had observed every covered change.

\noindent\textbf{Problem statement.}
Given task state $s$, candidate execution $a$, current approved outcome $A$, and a backend profile $p$ declaring $\partial_p$, persistent-outcome validation must establish $E$ and its completeness, classify the execution, and preserve that decision across failures. A workflow propagates success only for an execution with current approval and an \term{EXACT} result. Before durability, invalid approval, failed preconditions, and detected divergence cause rejection; after a durable or external boundary, divergent and indeterminate results remain visible and halt automatic continuation.

\noindent\textbf{Trust basis.}
Meaningful persistent-outcome validation separates target specification, outcome evidence, and decision recording. The task owner or upstream application supplies $A$, a registered backend observation interface establishes $E$, and the runtime compares them and persists the decision. Allowing the agent to derive $A$ from its own call or report $E$ would reduce validation to self-attestation. The approving party, runtime decision core, and registered observation interface therefore carry these respective responsibilities; the agent, proposed operation, and action-triggered backend logic are the objects under validation.

\section{Persistent-Outcome Validation Protocol}
\label{sec:design}

\begin{figure*}[t]
\centering
\includegraphics[width=\textwidth]{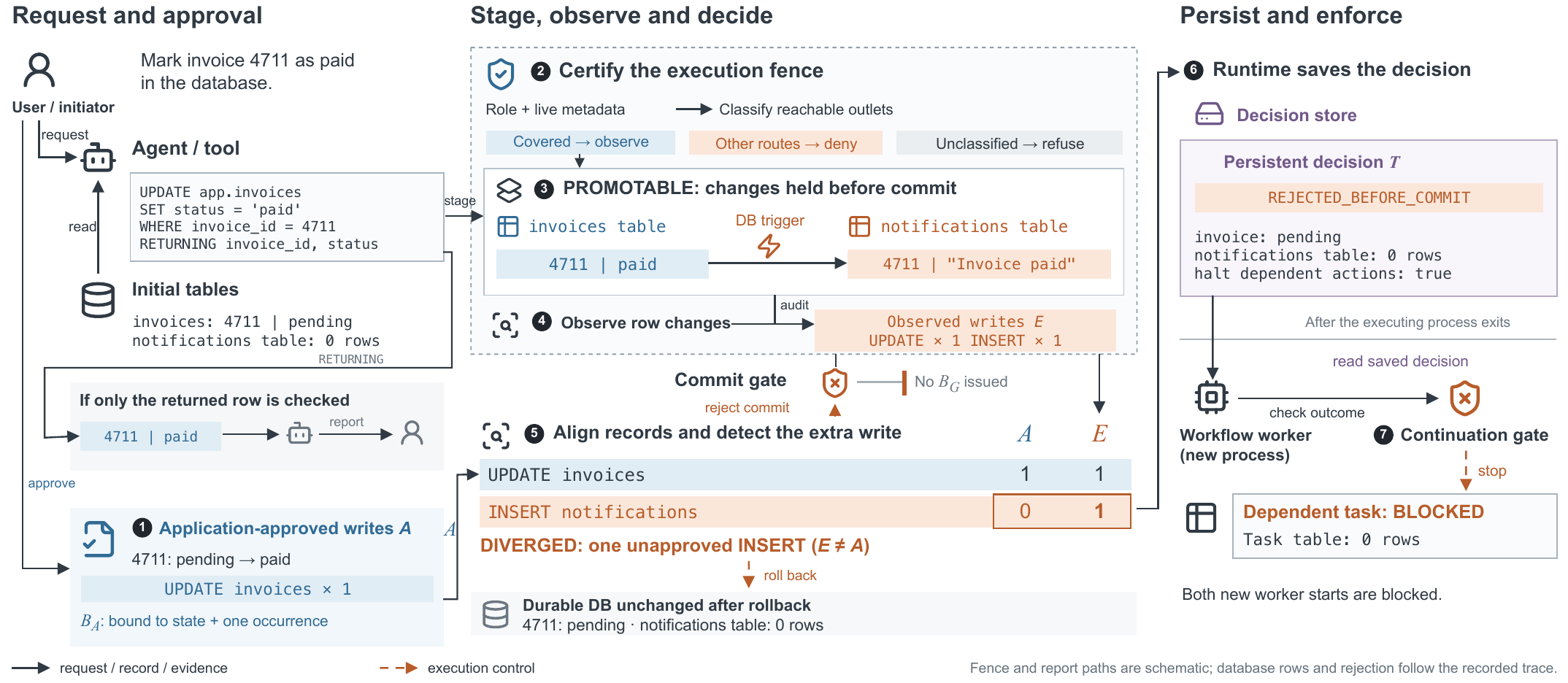}
\caption{\sys{} illustrated through invoice 4711 on a backend that holds changes for inspection before commit (Table~\ref{tab:protocol}). An unapproved trigger write causes rejection before commit; the saved decision blocks dependent work in later processes. A matching candidate receives a one-use commit permit $B_G$ after validation; none is issued in this rejected trace. Circular labels 1--7 locate the protocol steps.}
\label{fig:arch}
\end{figure*}

\sys{} implements the outcome classification in Section~\ref{sec:problem} through an \emph{Effect Commit Contract} that connects application approval, execution evidence, and continuation control for one execution occurrence. The protocol checks approval before execution, compares the resulting effects at the backend's control boundary, and saves the decision for subsequent workers. Figure~\ref{fig:arch} illustrates how these mechanisms cooperate on an invoice update.

\subsection{Protocol Overview}
\label{sec:overview}

Figure~\ref{fig:arch} follows the invoice update from application approval to rollback and a saved rejection. The three parts below correspond to the Introduction's problems; circular badges locate the protocol steps in the figure.

\noindent\gaplabel{1}\ Keep approval valid for the current execution. At \modulebadge{1}, the application approves changing invoice 4711 from \emph{pending} to \emph{paid}, with no notification-table insertion. The agent proposes the SQL command. Before executing it, the runtime binds approval to the current task state and one execution occurrence, providing the authority against which this attempt's effects will be checked. Section~\ref{sec:authority} describes the binding and its validation.

\noindent\gaplabel{2}\ Establish the actual effects before accepting the result. At \modulebadge{2}, the runtime checks the executing role and live backend configuration to certify that reachable routes to persistent effects are covered or blocked. An unclassified route prevents protected execution. Here, the scope covers both tables and the trigger path between them. At \modulebadge{3}, the command runs inside a private transaction held open for inspection. Both the invoice update and trigger insertion remain reversible. At \modulebadge{4}, transaction-local audit records expose the additional insertion alongside the requested update.

At \modulebadge{5}, the runtime compares observed effects $E$ with approved effects $A$ by operation, relation, and write count. Both contain one invoice update; only $E$ contains the notification insertion. The held candidate is \term{DIVERGED}, so the runtime rolls back both writes: the durable invoice remains \emph{pending}, and the notifications table remains empty. The command's successful return within the transaction therefore does not imply an accepted commit. Section~\ref{sec:establish} develops the observation and enforcement mechanisms for each backend capability.

\noindent\gaplabel{3}\ Preserve the decision across process boundaries. Later workers must learn of the rejection before acting on the command's apparent success. At \modulebadge{6}, the runtime records a pre-commit rejection and a halt decision. Here, \term{DIVERGED} classifies the candidate's mismatch; the saved rejection records its disposition. After the executing process exits, two fresh worker starts read this decision. The continuation gate at \modulebadge{7} blocks dependent execution, leaving the task table empty. Section~\ref{sec:recovery} explains the journal's role in recovery and continuation.

\subsection{Binding Approval to One Execution}
\label{sec:authority}

To close the approval-to-execution gap, \sys{} turns the approved content into a state-bound, one-use execution right through two steps: binding its context and claiming its occurrence.

\par\vspace{0.25em}
\noindent
\begingroup
\setlength{\tabcolsep}{0pt}
\setlength{\fboxsep}{2.5pt}
\centering
\begin{tabular}{@{}ccc@{}}
  \fcolorbox{rulegray}{white}{%
    \parbox[c][0.52in][c]{0.74in}{\centering
      \small\textbf{Approved }$A$\\[-1pt]
      \scriptsize invoice 4711\\[-1pt]pending $\rightarrow$ paid}}
  &
  \parbox[c]{1.72in}{\centering
    \begin{tabular}{@{}cc@{}}
      \parbox[c]{0.82in}{\centering\fontsize{7.0}{7.5}\selectfont 1) bind context}
        &
      \parbox[c]{0.82in}{\centering\fontsize{7.0}{7.5}\selectfont 2) claim occurrence}\\[-4pt]
      \multicolumn{2}{c}{\makebox[1.64in][c]{$\xrightarrow{\hspace{1.38in}}$}}\\[-4pt]
      \multicolumn{2}{c}{\parbox[c]{1.60in}{\centering\fontsize{6.2}{6.8}\selectfont
        mismatch / expired / claimed: stop}}
    \end{tabular}}
  &
  \fcolorbox{bindingblue}{bindingblue!7}{%
    \parbox[c][0.52in][c]{0.74in}{\centering
      \small\textcolor{bindingblue}{\boldmath$B_A$}\\[-1pt]
      \small\textbf{one-use right}\\[-1pt]
      \scriptsize occurrence claimed}}
\end{tabular}\par
\endgroup
\vspace{0.10em}

The application expresses the task requirements in an \emph{Effect Manifest}, specifying target objects, permitted changes, and state that must remain unchanged. Pre-execution state and business rules determine the concrete approved result. Our independent-approval experiments use hand-authored task specifications and separate business calculations to construct this content.

For example, a request to cancel unshipped items specifies an order and the cancellation scope. Business rules select eligible items from their initial status, determine their cancellation status, refund amount, and refund method, and update the order status accordingly. Shipped items and other business data remain unchanged. The same cancellation rule takes different orders and initial states as inputs to construct each task's approval. Once the user or application policy confirms the manifest, it becomes the reference for checking the agent's execution.

Let $M$ denote the confirmed manifest and $p$ the registered effect profile. The profile specifies a canonical representation for comparing approved and observed changes. We denote its normalization by $\operatorname{Canon}_p$ and construct
\begin{equation}
A := \operatorname{Canon}_p(M).
\label{eq:construct-approved}
\end{equation}
Profile $p$ fixes the comparison rules in advance. Normalization standardizes the representation of operations, targets, results, and multiplicities while preserving the targets and result constraints specified by the approval. The runtime renders the normalized content for approval and binds that same content in a signed license $L$, so the approving party and the execution check refer to the same outcome. Deployment frame $F$ fixes the matching profile, protected boundary, and trusted runtime configuration; validation of $L$ under $F$ establishes the authority of $A$ for the binding steps below.

The first step binds $A$ to the approving source, current task state, plan revision, adapter, policy, expiry, task instance, and transaction or request. We denote this current, one-use authority binding by $B_A$. A changed target, value, revision, state, adapter, or execution context invalidates the old approval. For example, if payment of invoice 4711 is approved and its vendor is then blocked, the changed task state prevents the old approval from authorizing that payment.

The second step atomically claims the occurrence assigned to this execution before the first effectful stage or irreversible dispatch. A stable approval identifier maps to one runtime bundle rather than minting another right on each presentation. Stale state and expiry fail before the claim; a competing worker, retry, or replay finds the occurrence already spent. A crash after the claim leaves it spent, so failure cannot recreate authority for a duplicate effect.

\subsection{Validating Persistent Outcomes at Backend Boundaries}
\label{sec:establish}
\label{sec:contract}
\label{sec:terminal}

Closing the second gap requires the runtime to establish the actual persistent outcome and turn its comparison with approval into a control decision at the backend boundary.

\begin{figure}[t]
\centering
\includegraphics[width=\columnwidth]{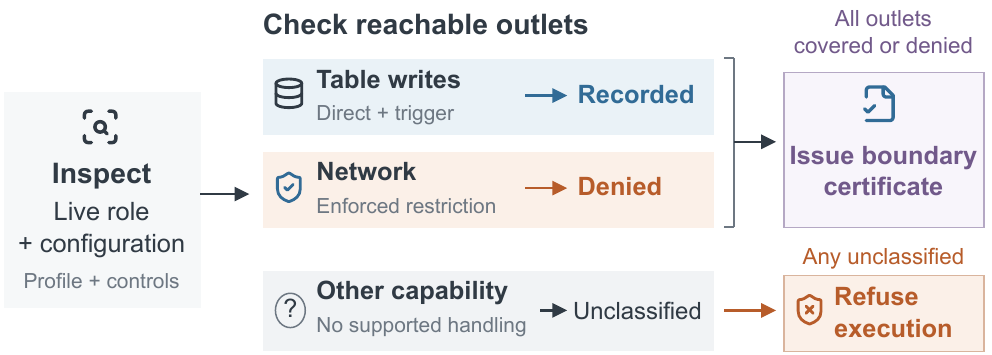}
\caption{Execution-fence certification: covered or denied outlets permit certification; any unclassified outlet prevents issuance.}
\label{fig:execution-fence}
\end{figure}

\paragraph{Certifying the execution boundary}
A profile declaration names the intended observation scope; certification checks whether the executor's reachable persistence and dispatch routes have corresponding controls. As Figure~\ref{fig:execution-fence} illustrates, the adapter combines the live execution identity and backend metadata with the host and network restrictions in force to classify these routes. Direct and trigger-induced writes to registered tables are covered by the observer; prohibited network access is blocked by enforcement; a route with no supported observation or blocking mechanism remains unclassified. The runtime issues a boundary certificate only when every reachable route has a supported disposition. An unclassified outlet prevents issuance and stops the operation before protected execution begins.

The PostgreSQL path certifies a preregistered, restricted deployment configuration. The adapter accepts its supported DML or fixed operation entry points and checks the registered relations, triggers, functions, and their permissions and configuration. The relation scope includes tables that supported trigger paths may modify, with protected write auditing on those relations. This scope is established by the deployment and includes relations that the current approval does not request to change.

A trusted metadata connection checks the live configuration. Unregistered functions, unsupported extensions, and reachable foreign-data or asynchronous-notification paths prevent admission; deployment isolation supplies the network restrictions outside the database. The certificate binds the outlet classifications and configuration, credential-provisioning, and enforcement digests to the backend, adapter, execution identity, and occurrence. The execution connection retains its restricted privileges. In the invoice example, the notification insertion therefore enters $E$ even though it is absent from $A$.

\paragraph{Establishing the actual outcome}
The records used to construct $E$ come from execution at the protected backend boundary. In the PostgreSQL path, \sys{} runs the action inside a rollback-capable private transaction and keeps that transaction open as the candidate. The registered relations are instrumented with write-recording triggers. A direct write and a write induced by an application trigger both produce audit records carrying the stage identifier and the affected relation and operation. Before collecting the records, the adapter executes \texttt{SET CONSTRAINTS ALL IMMEDIATE}, forcing deferred constraint triggers to run inside the held transaction. The observer can then collect their induced writes together with those already produced by the command.

For public database tasks using a net-state comparison, the adapter reads the registered data before execution and in the held candidate, then cross-checks their differences against this execution's write audit. These changes form $E$, which is compared with the approved outcome.

Let $R_a$ denote the change records collected and attributed to execution $a$. Using the same profile as the approved manifest, the observer constructs
\begin{equation}
E := \operatorname{Canon}_p(R_a).
\label{eq:construct-observed}
\end{equation}
The manifest and backend records may have different input formats; $\operatorname{Canon}_p$ denotes their respective mappings into the profile's common effect representation. The mapping preserves repeated effects and the comparison fields required by the profile. In Figure~\ref{fig:arch}, the command returns invoice 4711 as \emph{paid}, while the audit records contain both the invoice update and the notification insertion. The relation-write illustration shows their operations, relations, and multiplicities, which expose the extra insertion. Value-sensitive profiles additionally retain the result values or value digests in their comparison keys. For the held transaction, $E$ describes candidate changes awaiting commit, so detecting a mismatch still leaves both writes reversible.

Constructing $E$ organizes the available records; establishing $\mathrm{Complete}_p(E)$ determines whether they cover the execution within the registered boundary. The certified profile fixes $\partial_p$ before execution by registering the state objects, result-propagation paths, and evidence channels it covers and by ruling out unclassified outlets for that executor. The observer must also establish the required evidence coverage and attribute the records to this execution. In Figure~\ref{fig:arch}, the invoice and notification relations, the direct and trigger paths, and their held-transaction changes belong to the same observation scope. Their records therefore enter one $E$. Boundary certification and execution evidence jointly support completeness; a missing required channel or unresolved attribution leaves the result indeterminate.

Once completeness is established, an extra element in $E-A$, a missing element in $A-E$, or a changed result represented by the profile yields divergence. The PostgreSQL runtime locks the registered relations before reading the initial state and holds these locks until commit or rollback, stabilizing the data used in the comparison. Administrative role grants, function implementations, and external isolation are deployment-managed configuration; this path requires them to remain stable throughout protected execution. The final gate reads fresh metadata and compares the resulting certificate with the one bound to the permit, rejecting changes visible at that check. It also rechecks transaction identity, candidate state, and audit records before committing the same transaction. A failed check rolls back the held transaction. Non-transactional sequence residue is bounded separately and is not included in a zero-relational-residue claim.

\paragraph{Enforcing the comparison}
Once $E$ has been established, backend capability determines whether a mismatch can be rejected before durability or must be classified and contained afterward. Every mutating adapter declares one capability class, and admission checks that declaration against the interface the adapter actually controls. All classes claim the authorized occurrences before the first effectful stage or dispatch. Immediately before promotion or dispatch, the adapter revalidates state, authority, and its own identity, then consumes a permit bound to that particular stage or request. The capability class determines what can be checked before this gate and what can only be learned afterward.

Some backends let \sys{} hold and observe the same candidate that it later promotes; we call this capability \term{PROMOTABLE}. The runtime controls the write credential or commit entry point, revalidates state and authority, and requires complete observation before promotion. A mismatch is discarded before durability; an equal candidate receives a stage-bound $B_G$ and may be promoted. Database transactions, atomic file replacement, and server-side draft/promote interfaces can provide this shape.

Some remote backends cannot stage a candidate for later promotion but support a request path exclusively controlled by \sys{}; we call this capability \term{MEDIATED}. Adapter admission requires an exclusive credential path, deterministic request generation from the approved manifest, and an idempotency mechanism that prevents repeated requests from duplicating effects. The adapter constructs the request from $A$, binds it in $B_G$, and consumes the permit before dispatch. The agent or an ordinary wrapper can therefore neither bypass the gateway nor add unapproved request fields. Post-dispatch evidence supports exact, diverged, or indeterminate classification under the same completeness requirements.

For external calls that neither support staging nor meet the full gateway-control conditions above, we use the \term{OPAQUE} path. The adapter constrains requests sent through it and binds them to approval, then classifies the outcome using available post-execution evidence. Complete evidence matching approval permits continuation; a complete mismatch records divergence; insufficient evidence records uncertainty. The latter two outcomes block dependent execution, and this path provides no rollback of effects already realized.

Table~\ref{tab:protocol} connects effect classification to execution outcomes. An exact candidate is recorded as successfully committed only after promotion is confirmed. Remote readback establishes the current state of registered objects; recording an exact execution outcome additionally requires matching approval and satisfying the coverage and attribution conditions of Definition~2. A failed check before commit or dispatch instead produces rejection, as in Figure~\ref{fig:arch}. After the operation may have taken effect, the journal records divergence for a complete mismatch or uncertainty for an unresolved outcome. The journal thus records how execution ended, while the effect classification describes whether the observed candidate or outcome matched approval.

\begin{table}[t]
\centering
\caption{Capability-specific control points. Every path claims its assigned authorization occurrences before staging or dispatch; a later permit binds the checked stage or request.}
\label{tab:protocol}
\footnotesize
\begin{tabular}{@{}p{1.9cm}p{2.1cm}p{3.4cm}@{}}
\toprule
\textbf{Class} & \textbf{Control point} & \textbf{Supported outcome} \\
\midrule
\term{PROMOTABLE} & inspect a held candidate before durability & mismatch or incomplete evidence: reject; confirmed equal promotion: exact; commit issued but result unestablished: indeterminate \\
\addlinespace[2pt]
\term{MEDIATED} & exclusive gateway; manifest-derived requests and idempotency; read back afterward & complete and equal: exact; complete and unequal: diverged; incomplete evidence or unresolved response loss: indeterminate \\
\addlinespace[2pt]
\term{OPAQUE} & constrain requests through the adapter; inspect available evidence afterward & complete and equal: exact; complete mismatch: diverged; incomplete evidence: indeterminate; no rollback of realized effects \\
\bottomrule
\end{tabular}
\end{table}

Every backend path must preserve its decision for later workers, so that rejection, divergence, or uncertainty continues to block dependent execution after the detecting process exits.

\subsection{Persistent Recovery and Continuation}
\label{sec:recovery}

The capability-specific paths pass their execution decisions to later processes through the terminal journal. If the process exits before a database candidate commits, the uncommitted transaction rolls back when its connection terminates. An exit after business-state persistence but before terminal persistence leaves an execution requiring reconciliation. A new worker consults the existing terminal to decide whether work may proceed; a missing terminal keeps continuation blocked pending reconciliation. An occurrence already claimed remains spent.

The journal is also the release point for continuation. Schedulers, queue consumers, and recovery workers consult it before dispatching a dependent action. Delayed work remains part of the same occurrence only when a registered continuation presents the matching occurrence identifier, continuation identifier, and authority lineage, then records its terminal receipt before releasing dependent work. An independent reader, mismatched lineage, or replay does not inherit the original authority. Dependent actions run only after reading a persistent terminal confirming exact completion; a missing terminal or a rejected, diverged, or indeterminate outcome keeps continuation blocked pending resolution. A spent occurrence cannot authorize an automatic resend. A later compensation or retry is a new business effect and requires new application authority. At task scope, \sys{} preflights known capability and preconditions before the first external write, preserves exact terminals for actions already completed, and reports which remaining effects require resolution.

Let $T$ denote the journal record for this execution. Recording confirmed exact completion requires agreement among the approved outcome, observed outcome, authority claim, and final-gate permit:
\begin{equation}
\begin{array}{c}
\text{$T$ records confirmed exact completion}\;\Longrightarrow\\[-1pt]
\mathrm{EffectExact}_p(A,E)\\[-1pt]
{}\land\mathrm{AuthValid}(B_A,t_c)\\[-1pt]
{}\land\mathrm{GateValid}(B_G,t_g).
\end{array}
\label{eq:assurance}
\end{equation}
$\mathrm{AuthValid}$ records the successful current-state claim of one occurrence at $t_c$, while $\mathrm{GateValid}$ records that the checked stage or request consumed its bound permit at $t_g$. Both describe completed checks, not reusable credentials. The permit must be consumed to complete the authorized transition; requiring it to remain unspent would make a successful terminal self-contradictory.

Algorithm~\ref{alg:effect-assured-execution} consolidates these three stages into one capability-aware execution path. For a profile requiring outlet certification, \textsc{CheckBoundary} obtains a certificate before staging and \textsc{ValidateAndConsume} recaptures it at the final gate, together with the live-stage or request checks; other profiles use their declared admission checks. Every return passes through \textsc{Finish}, which durably writes $T$ before releasing an exact execution's dependents or blocking continuation for any other result.

\begingroup
\algtext*{EndIf}
\begin{algorithm}[t]
\caption{Capability-aware execution of the Effect Commit Contract.}
\label{alg:effect-assured-execution}
\footnotesize
\begin{algorithmic}[1]
\Require action $a$, state $s$, frame $F$, signed license $L$, profile $p$, capability $c$
\Ensure on normal return, persistent terminal $T$ and continuation decision
\Statex \hspace{\algorithmicindent}\textit{Authority gate}
\State $(A,B_A,\mathit{authOK},t_c) \gets \Call{ValidateAndClaim}{a,s,F,L,c}$
\State $(E,B_G,t_g) \gets (\bot,\bot,\bot)$; $\mathit{outcome} \gets \bot$
\If{$\neg \mathit{authOK}$}
    \State \Return \Call{Finish}{\textsc{Reject}}
\EndIf
\State $(\mathit{boundaryOK},\mathit{cert}) \gets \Call{CheckBoundary}{p,a}$
\If{$\neg\mathit{boundaryOK}$}
    \State \Return \Call{Finish}{\textsc{Reject}}
\EndIf
\If{$c=\term{PROMOTABLE}$}
    \Statex \hspace{\algorithmicindent}\textit{Promotable: inspect before durability}
    \State $H \gets \Call{Stage}{a,s}$; $(E,\mathit{complete}) \gets \Call{Observe}{H}$
    \If{$\neg \mathit{complete}$ \textbf{ or } $E\neq A$}
        \State \Call{Discard}{H}; \Return \Call{Finish}{\textsc{Reject}}
    \EndIf
    \State $B_G \gets \Call{BindStagePermit}{B_A,H,E,\mathit{cert}}$
    \Statex \hspace{\algorithmicindent}\textit{Recheck boundary and candidate}
    \State $(\mathit{gateOK},t_g) \gets \Call{ValidateAndConsume}{B_G,p}$
    \If{$\neg \mathit{gateOK}$}
        \State \Call{Discard}{H}; \Return \Call{Finish}{\textsc{Reject}}
    \EndIf
    \State $r\gets\Call{Promote}{H}$; $\mathit{outcome}\gets\Call{PromotionOutcome}{r}$
\Else
    \Statex \hspace{\algorithmicindent}\textit{Remote: constrain before dispatch; classify after}
    \State $q \gets \Call{RegenerateOrConstrain}{A,c}$
    \State $B_G \gets \Call{BindRequestPermit}{B_A,q,A,\mathit{cert}}$
    \State $(\mathit{gateOK},t_g) \gets \Call{ValidateAndConsume}{B_G,p}$
    \If{$\neg \mathit{gateOK}$}
        \State \Return \Call{Finish}{\textsc{Reject}}
    \EndIf
    \State $(r,\mathit{mayHaveSent}) \gets \Call{Dispatch}{q}$
    \If{$\neg\mathit{mayHaveSent}$}
        \State \Return \Call{Finish}{\textsc{Reject}}
    \EndIf
    \State $(E,\mathit{complete}) \gets \Call{ReadBack}{c,r}$
    \State $\mathit{outcome}\gets\Call{ClassifyReadback}{E,\mathit{complete},A}$
\EndIf
\Statex \hspace{\algorithmicindent}\textit{Persistent terminal before continuation}
\State \Return \Call{Finish}{$\mathit{outcome}$}
\end{algorithmic}
\end{algorithm}
\endgroup

\subsection{Prototype Realization}
\label{sec:prototype}

The prototype is a Python runtime with registered adapters, effect profiles, and observers. Its mediated HTTP adapter attaches idempotency keys and available state preconditions to manifest-derived requests. In the STATE-Bench live-agent integration, tool calls first modify a private copy of the task state; the runtime then checks the candidate and decides whether to save it. When the model finishes normally, the integration layer triggers any outstanding finalization check.

The implementation persists the approval-to-bundle issuance ledger, occurrence ledger, terminal journal, and divergence receipts. A stable source nonce lets the issuance ledger deduplicate each issuer's approval, while the occurrence ledger records its one-time use.

Schema-closed approval, bundle, and permit records are HMAC-authenticated under separate keys; content-addressed contract and observer digests identify the code and profile used for each decision. The trusted boundary contains the runtime, authority-key holder, approval channel, policy store, and registered adapters and observers.

\section{Evaluation}
\label{sec:eval}

We evaluate \sys{} primarily on public database tasks and shopping, customer-support, and travel tasks. Controlled comparisons and ablations test outcome checking and execution binding, while live-agent experiments assess task completion across agent configurations. Further experiments examine continuation after process failures, cross-backend behavior, and runtime overhead. The evaluation addresses five questions:

\begin{itemize}
\item[\textbf{RQ1}] Compared with call-level rules and conventional workflows, can \sys{} detect outcome mismatches and prevent incorrect commits?
\item[\textbf{RQ2}] Which \sys{} components are necessary?
\item[\textbf{RQ3}] Can \sys{} preserve successful task completion and integrate with different live-agent configurations?
\item[\textbf{RQ4}] Do persistent decisions correctly govern continuation after failures and process restarts?
\item[\textbf{RQ5}] How does \sys{} behave across backends and deployment conditions, and what is its runtime overhead?
\end{itemize}

\begin{table*}[t]
\centering
\caption{Main experimental populations and repetitions. Execution counts include all conditions and variants within each block.}
\label{tab:breadth}
\small
\setlength{\tabcolsep}{5pt}
\begin{tabularx}{\textwidth}{@{}>{\raggedright\arraybackslash}X>{\raggedright\arraybackslash}p{3.3cm}>{\raggedright\arraybackslash}p{3.2cm}r@{}}
\toprule
\textbf{Experiment block} & \textbf{Distinct tasks / cases} & \textbf{Repetitions per cell} & \textbf{Executions} \\
\midrule
PostgreSQL: independent approval & 3 public tasks & 3 & 198 \\
AgentSpec: visible / hidden writes & same 3 public tasks & 3 & 81 \\
STATE: independent approval & 12 public tasks & 3 & 828 \\
STATE: clean-reference comparison & 206 public tasks & 3 & 7,416 \\
Live-agent integration & 6 public tasks & 3 & 108 \\
Cross-process recovery & 18 public tasks & 5 & 540 \\
Runtime cost & same 18 public tasks & 30 pairs per task & 1,080 \\
ToolSandbox & 4 public scenarios & 3 & 72 \\
MySQL & 8 constructed cases & 5 & 240 \\
GitHub & 8 operation cases & 5 & 120 \\
\bottomrule
\end{tabularx}
\par\smallskip
\begin{minipage}{\textwidth}
\footnotesize
A cell fixes the task or case, execution condition, and variant. Task sets overlap across rows. Recovery additionally starts 1,080 consumer processes. Certification, concurrency, and targeted reruns are specified with their respective results.
\end{minipage}
\end{table*}

\subsection{Experimental Setup}
\label{sec:setup}

\emph{Comparison conditions.} Public-task comparisons use direct execution, conventional workflow controls, and AgentSpec call-level rules~\cite{wang2026agentspec}. Direct PostgreSQL execution also uses transactions; the AgentSpec condition uses its original rule interpreter with our SQL rule to check visible write operations and target relations. The STATE-Bench workflow condition executes tool calls in a private copy and discards that copy on a tool error. Ablations remove specified \sys{} checks or bindings under the same task and fault conditions. Mechanism diagnostics additionally use \emph{Complete-Effect Audit}, an instrumented complete-observation ceiling, which is reported separately from deployable controls.

\emph{Tasks and approval sources.} The main experiments use database tasks from MCPMark\footnote{\url{https://github.com/eval-sys/mcpmark}} and shopping, customer-support, and travel tasks from STATE-Bench\footnote{\url{https://github.com/microsoft/STATE-Bench}}. For three selected database tasks and twelve business tasks, we independently construct approved outcomes before execution from public requests, initial state, and business rules. A broader comparison on 206 business tasks uses previously saved clean-execution results as targets to test checking with a supplied reference. ToolSandbox\footnote{\url{https://github.com/apple/ToolSandbox}} contributes four public scenarios; MySQL and GitHub use predeclared operation cases. Table~\ref{tab:breadth} reports distinct tasks or cases, repetitions, and execution counts.

\emph{Execution and environment.} Controlled experiments execute fixed SQL or replay public tool calls. Live-agent experiments generate independent call sequences with three configurations: a minimal DeepSeek API loop and DSH~0.1.5-rc.2, both using \texttt{deepseek-flash} with thinking disabled, and Codex CLI~0.154.0 using GPT-5.6 Sol with high reasoning. Each task supplies the same public request bundle and business-tool schemas to all three configurations, under direct and \sys{}-integrated execution. Six tasks, three configurations, two conditions, and three repetitions yield 108 formal episodes. Business tools run in local simulators, while model calls use the actual APIs or CLI. Local experiments used an Apple M5 host and Python~3.12.13.

\emph{Outcome measurement.} We measure normal task completion and incorrect persistent changes from the saved business state, and check continuation decisions through fresh processes that read persisted terminal records. Native benchmark scores provide post-execution assessment. Each condition decides using its assigned runtime evidence; raw calls, states, and decisions are retained in the experimental artifacts. The predeclared live-agent shopping comparison matches newly created cart items by their business fields and references, allowing their system-assigned identifiers to differ while preserving existing item identifiers. Cart-item list order is ignored; item multiplicities and the remaining business state are still checked.

\begin{table}[t]
\centering
\caption{Public-task comparisons. Clean rows count correct commits ($\uparrow$); fault rows count incorrect commits ($\downarrow$). Each task/variant is repeated three times.}
\label{tab:public-outcomes}
\small
\setlength{\tabcolsep}{3pt}
\begin{tabularx}{\columnwidth}{@{}>{\raggedright\arraybackslash}Xrrr@{}}
\toprule
\multicolumn{4}{@{}l}{\textbf{(a) PostgreSQL: 3 independently approved tasks}} \\
\midrule
\textbf{Outcome} & \textbf{Direct} & \textbf{AgentSpec} & \textbf{\sys{}} \\
\midrule
Clean $\uparrow$ & 9/9 & 9/9 & 9/9 \\
Hidden extra write $\downarrow$ & 9/9 & 9/9 & 0/9 \\
Missing row $\downarrow$ & 9/9 & 9/9 & 0/9 \\
Altered value $\downarrow$ & 9/9 & 9/9 & 0/9 \\
\addlinespace[5pt]
\multicolumn{4}{@{}l}{\textbf{(b) STATE: 12 independently approved tasks}} \\
\midrule
\textbf{Outcome} & \textbf{Direct} & \textbf{Error-reject} & \textbf{\sys{}} \\
\midrule
Clean $\uparrow$ & 36/36 & 36/36 & 36/36 \\
Omit write $\downarrow$ & 36/36 & 36/36 & 0/36 \\
Duplicate write $\downarrow$ & 9/36 & 9/36 & 0/36 \\
Alter value $\downarrow$ & 36/36 & 36/36 & 0/36 \\
Extra field change $\downarrow$ & 36/36 & 36/36 & 0/36 \\
\addlinespace[5pt]
\multicolumn{4}{@{}l}{\textbf{(c) STATE: 206 tasks with clean references}} \\
\midrule
\textbf{Outcome} & \textbf{Direct} & \textbf{Error-reject} & \textbf{\sys{}} \\
\midrule
Clean $\uparrow$ & 618/618 & 597/618 & 618/618 \\
Omit write $\downarrow$ & 354/618 & 138/618 & 0/618 \\
Duplicate write $\downarrow$ & 138/618 & 138/618 & 0/618 \\
\bottomrule
\end{tabularx}
\par\smallskip
\begin{minipage}{\columnwidth}
\footnotesize
Error-reject discards the private candidate on a tool error. Faults that leave the approved state unchanged remain in the denominators.
\end{minipage}
\end{table}

\subsection{RQ1: Can Outcome Checking Prevent Incorrect Commits?}
\label{sec:rq1}

\begin{figure}[t]
\centering
\includegraphics[width=\columnwidth]{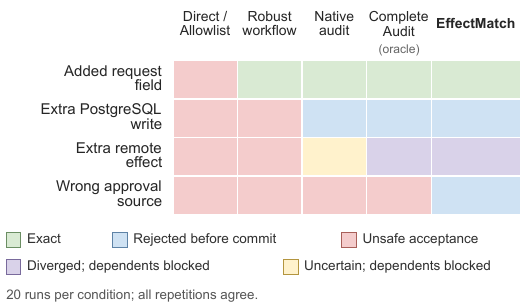}
\caption{Four mechanism scenarios from the full matrix: 20 runs per condition with identical within-cell outcomes. Direct and Allowlist ran separately and share a column. Complete Audit is the instrumented observation oracle. Diverged and uncertain outcomes block dependents without undoing the remote effect.}
\label{fig:comparators}
\end{figure}

An allowed call can still produce a result that violates approval. RQ1 compares call-level rules, conventional workflow controls, and \sys{} on public tasks, asking whether they preserve correct execution while preventing incorrect commits. Table~\ref{tab:public-outcomes} first reports independently approved database and business tasks, then extends the comparison to 206 tasks with saved clean-execution references. Figure~\ref{fig:comparators} uses separate mechanism cases to explain the differences between evidence levels.

The three database tasks cover customer migration, employee hierarchy management, and sales/music charts. AgentSpec's SQL rule checks visible write operations and target relations: in a separate explicit-write control, it rejects all nine unapproved writes. However, when the submitted SQL stays unchanged and fixture triggers introduce an extra write, a missing row, or an altered value, the rule has no resulting-row evidence on which to reject. \sys{} compares candidate changes with the independently specified approval before commit and rejects all 27 mismatching executions, preserving the initial relation data. All conditions complete the clean cases. The distinction is whether the transaction should commit, rather than whether execution uses a transaction.

The twelve business tasks separate tool status from business correctness. Omission and duplication modify native calls; altered values and extra field changes are injected into the candidate snapshot after replay. These changes can leave an incorrect state without a tool error, so error-triggered rollback still accepts them. Duplication also exposes unnecessary rejection: nine executions change the required outcome, while 27 still reach the approved state. \sys{} rejects the nine mismatches and commits the 27 matching candidates. Error-reject accepts all nine mismatches yet blocks twelve matching candidates across four tasks. Across the four fault types, \sys{} rejects all 117 mismatches, representing 39 task/variant cases repeated three times. Checking the realized state separates harmful changes from calls whose repetition leaves the result intact.

\begin{figure}[t]
\centering
\includegraphics[width=0.65\columnwidth]{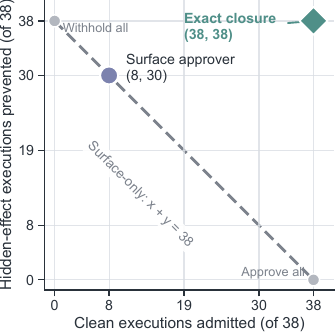}
\caption{Safety--utility boundary over 38 frozen STATE-Bench pairs. GPT-5.6 Sol (ultra) reaches $(8,30)$; exact post-execution closure reaches $(38,38)$.}
\label{fig:information-boundary}
\end{figure}

The broader clean-reference comparison preserves this distinction. Error-reject reduces incorrect omission commits but misses every state-changing duplication in Table~\ref{tab:public-outcomes}(c). It also withholds 21 correct clean outcomes because their trajectories recover from an intermediate tool error. \sys{} retains all 618 clean outcomes and rejects the mismatching candidates. Ineffective omissions and duplications remain eligible for commit because their final state still matches the reference.

The mechanism diagnostics distinguish what each source of evidence adds. Here, Direct trusts the API response; Allowlist adds parameter, schema, and resource checks. Robust Workflow adds request constraints, state preconditions, readback, and transaction/retry controls; NativeAudit additionally uses ordinary backend audit evidence. Complete-Effect Audit receives the complete protected-effect set and a locally derived expected set. All six conditions complete the clean control in 20/20 runs each; Figure~\ref{fig:comparators} displays four differentiating scenarios. The full eight-scenario matrix remains in the experimental artifacts.

Robust Workflow prevents the extra request field by constraining dispatch. NativeAudit additionally exposes the hidden PostgreSQL write and enables rejection before commit. For an already-realized remote effect, NativeAudit returns uncertainty, while complete observation establishes divergence; both stop dependent work. Finally, substituting the approval source leaves the observed effects unchanged, so even Complete-Effect Audit accepts the local match. \sys{} rejects the untrusted source. Given the same approved target and complete observation, an in-transaction postcondition check can also reject the extra write. These comparisons show how outcome evidence changes commit decisions. RQ2 then examines whether approval still belongs to the current execution, whether the committed object is the checked candidate, and whether the saved decision continues to govern later work.

Figure~\ref{fig:information-boundary} illustrates the corresponding information limit on 38 clean/hidden-effect pairs with identical pre-execution packets. The surface approver evaluates these packets in one tool-free batch; its six denials and 24 escalations both withhold execution. The same surface decision applies to both members of each pair: admitting $x$ clean cases prevents at most $38-x$ hidden counterparts. Post-execution observation distinguishes the pairs and reaches $(38,38)$.

\begin{table}[t]
\centering
\caption{Paired ablations grouped by mechanism. Public task/variant pairs have three repetitions; dedicated cases have 20. Outcomes report counts for full execution $\rightarrow$ removal.}
\label{tab:ablation}
\small
\setlength{\tabcolsep}{3pt}
\begin{tabularx}{\columnwidth}{@{}>{\raggedright\arraybackslash}p{2.05cm}>{\raggedright\arraybackslash}p{1.65cm}>{\raggedright\arraybackslash}X@{}}
\toprule
\textbf{Removed check} & \textbf{Case} & \textbf{Full $\rightarrow$ removal} \\
\midrule
Result comparison & PG tasks & 27 reject $\rightarrow$ 27 wrong commits \\
 & STATE tasks & 117 reject $\rightarrow$ 117 wrong commits \\
\addlinespace[3pt]
Complete $E$ & Extra write & 20 reject $\rightarrow$ 20 extra-write commits \\
\addlinespace[3pt]
Approval source ($B_A$) & Source swap & 20 reject $\rightarrow$ 20 wrong-source approvals accepted \\
\addlinespace[3pt]
Execution binding & Replay & 20 reject $\rightarrow$ 20 duplicate effects \\
 & PG occurrence & 9 reject $\rightarrow$ 9 wrong-instance commits \\
 & STATE occurrence & 36 reject $\rightarrow$ 36 wrong-instance commits \\
\addlinespace[3pt]
Durable hazard ($T$) & Continuation & 20 block $\rightarrow$ 20 dependent actions \\
\addlinespace[3pt]
Candidate binding ($B_G$) & Replacement & 20 reject $\rightarrow$ 20 wrong-candidate commits \\
\addlinespace[3pt]
Final boundary check & Certificate reuse & 20 reject $\rightarrow$ 20 stale-boundary commits \\
 & PG joint recheck & 9 reject $\rightarrow$ 9 stale-boundary commits \\
\bottomrule
\end{tabularx}
\par\smallskip
\begin{minipage}{\columnwidth}
\footnotesize
PG and STATE use three and twelve public tasks. Occurrence cases remove occurrence/pre-state binding. The two boundary rows remove fresh certificate capture and the combined boundary/audit recheck, respectively.
\end{minipage}
\end{table}

\subsection{RQ2: Which Components Are Necessary?}
\label{sec:rq2}

RQ1 showed how outcome checking changes commit decisions. RQ2 removes individual checks or related binding checks to identify where incorrect results or invalid approvals are admitted. Public-task ablations test result comparison, execution binding, and final rechecking; targeted cases isolate complete observation, approval provenance, persistent terminals, and candidate binding. Table~\ref{tab:ablation} pairs full \sys{} with each removal.

Disabling result comparison commits all 27 incorrect database candidates and all 117 incorrect business-task candidates, while full \sys{} rejects them. The other execution conditions remain unchanged, isolating whether the realized result is checked against approval. The observation ablation separately tests the comparison's input: omitting the additional write from the observed effects makes an incorrect result appear to match approval. The provenance case tests the target instead. Its effects match a locally supplied target, but that target comes from an untrusted source. Full \sys{} rejects it through the approval-source check in $B_A$.

Execution binding checks whether approval belongs to the current occurrence. Across three database and twelve business tasks, the candidate business result remains correct while the occurrence identifier changes. Full \sys{} rejects nine and 36 executions, respectively; removing occurrence and pre-state binding admits all of them. Matching results and valid approval ownership are therefore independent conditions. The replay case shows the consequence of reuse: removing one-use enforcement produces an additional external effect in every repetition. The persistent terminal $T$ carries the decision into subsequent execution. Without its durable hazard record, previously detected divergence is lost and a dependent action is dispatched. Full \sys{} rejects replay and preserves the divergence that blocks continuation.

Candidate binding ensures that the committed object is the one already checked. In the file-backed invoice case, the runtime checks a \emph{paid} candidate and issues a valid permit; a second candidate for the same action then carries \emph{cancelled}. Removing the $B_G$ candidate-binding checks at both runtime and adapter lets that permit commit the replacement, whereas full \sys{} rejects all substitutions. Signature, expiry, current policy, base-state validation, and one-use enforcement remain active. Candidate binding ties the validation decision to the specific result selected for commit.

Final rechecking also establishes whether the execution environment still satisfies certification. The database case grants temporary-table permission after staging, changing the permission boundary while leaving the protected relation catalog unchanged. Reusing the initial boundary certificate admits commit; fresh certification detects the change and rejects. The retained catalog check does not cover this database-level privilege. Committed rows still match approval: the failure is admission after certification conditions cease to hold. Across the three public database tasks, disabling the final boundary/audit recheck likewise admits all nine post-authorization deployment changes. That ablation removes the combined final recheck, while the targeted case isolates fresh certificate capture. The candidate-binding and certificate-capture studies both commit every clean control (20 per condition), and follow-up permit replays produce no second commit.

\subsection{RQ3: Live-Agent Task Completion and Integration}
\label{sec:rq3}

The preceding questions examined how incorrect results are blocked and why the individual checks are necessary. RQ3 asks whether outcome checking can integrate with live agents while preserving successful task completion. We select six public STATE tasks covering shopping-cart changes, order cancellations, and travel-booking updates, and run DeepSeek API, DeepSeek+DSH, and Codex configurations. Each task has three repetitions under both direct execution and \sys{} integration, yielding 108 independently generated live episodes. Business tools operate local simulators, and approval targets are independently established before model execution.

All 108 episodes produce candidate business states that match their approval targets. Table~\ref{tab:live-agent-completion} reports whether the correct result is durably saved. Direct execution completes throughout; integrated DSH and Codex also complete throughout. For DeepSeek API, correcting finalization handling and rerunning the affected episodes produces correctly saved results for the remaining task repetitions.

These tasks combine required changes with state that must be preserved. The shopping task permits gift wrapping only for the specified product; the order task must retain already-shipped items; and the travel task must preserve booking preferences when changing a flight. Correct completion therefore requires both the requested changes and preservation of unaffected state. The observed candidates satisfy these requirements, and \sys{}'s checks retain the corresponding paths to successful execution.

\begin{table}[t]
\centering
\caption{Correct durable outcomes on six live-agent tasks, with three repetitions per configuration and condition.}
\label{tab:live-agent-completion}
\small
\setlength{\tabcolsep}{5pt}
\begin{tabularx}{\columnwidth}{@{}>{\raggedright\arraybackslash}X>{\centering\arraybackslash}p{1.8cm}>{\centering\arraybackslash}p{2.2cm}@{}}
\toprule
\textbf{Agent configuration} & \textbf{Direct} & \textbf{With \sys{}} \\
\midrule
DeepSeek API & 18/18 & 18/18$^{*}$ \\
DeepSeek+DSH & 18/18 & 18/18 \\
Codex & 18/18 & 18/18 \\
\bottomrule
\end{tabularx}
\par\smallskip
\begin{minipage}{\columnwidth}
\footnotesize
$^{*}$Original integrated runs completed 51/54. The marked cell combines 15 original successes with three successful targeted reruns after host-triggered finalization was added at normal agent completion.
\end{minipage}
\end{table}

The three configurations organize tool execution through a minimal API loop, the DSH harness, and Codex CLI, respectively. \sys{} follows the same integration principle in each: business tools first form a candidate state, and the runtime compares it with independently approved targets before saving it. This places outcome checking at the business-state persistence point across different agent frameworks, without reading the model's internal reasoning. The results support the feasibility of this integration on the tested tasks and configurations.

\subsection{RQ4: Continuation after Failures and Process Restarts}
\label{sec:rq4}

RQ3 examined successful task completion with live agents. RQ4 asks whether a new worker can correctly decide what may continue after the executing process exits. We select six public multistep tasks from each of shopping, customer support, and travel, replay their tool calls, and use previously saved clean-execution results as comparison targets. Six normal or interrupted scenarios, each repeated five times per task, produce 540 execution processes. After each exits, two fresh processes read the saved business state and terminal records, yielding 1,080 continuation checks.

Table~\ref{tab:process-recovery} shows how the exit point affects persistence and continuation. Normal completion and exit after terminal persistence both permit continuation; exit while the candidate remains staged preserves the initial state and stops subsequent work. No case releases work without a valid result, and the two fresh readers produce no duplicate acknowledgement records.

\begin{table}[t]
\centering
\caption{Persistence and continuation after process exit. Each row counts 90 executions: 18 tasks with five repetitions.}
\label{tab:process-recovery}
\small
\setlength{\tabcolsep}{3pt}
\begin{tabularx}{\columnwidth}{@{}>{\raggedright\arraybackslash}X>{\centering\arraybackslash}p{1.15cm}>{\centering\arraybackslash}p{1.3cm}>{\centering\arraybackslash}p{1.15cm}@{}}
\toprule
\textbf{Scenario} & \textbf{Correct state} & \textbf{Terminal} & \textbf{Release} \\
\midrule
Normal completion & 90 & 90 & 90 \\
Exit after staging & 0 & 0 & 0 \\
Exit after state save, before terminal & 90 & 0 & 0 \\
Exit after terminal save & 90 & 90 & 90 \\
Omit first write; exit after staging & 0 & 0 & 0 \\
Omit first write; exit after terminal save & 35 & 90 & 35 \\
\bottomrule
\end{tabularx}
\par\smallskip
\begin{minipage}{\columnwidth}
\footnotesize
Correct state counts saved results matching the reference. Terminal counts include success and rejection. Release uses a harness acknowledgement; two fresh readers produce at most one saved acknowledgement per execution.
\end{minipage}
\end{table}

Recovery checks both the successful terminal and its bound durable state. A saved success therefore survives the producing process, and a new worker can use it to continue. The interval between business-state persistence and terminal persistence exposes a gap: all 90 executions leave the correct state, but continuation is withheld because the terminal is missing. The new workers in this experiment read existing terminals without reconstructing missing records, so these executions await reconciliation before continuation. Release is represented by a unique acknowledgement record in the harness.

The omitted-write results show that continuation depends on the realized state. In 35 of 90 executions, omitting the first write leaves the final result unchanged, so checking permits persistence and continuation. The remaining 55 candidates mismatch the target; the runtime saves rejection, and subsequent workers stop. This carries forward the outcome rule from RQ1 and RQ2: a changed execution trace need not imply an incorrect business result.

The independently approved PostgreSQL customer-migration task exhibits the same recovery behavior. Five repetitions at each exit point yield 0/5 continuation acknowledgements before commit, 0/5 after commit but before terminal persistence, and 5/5 after terminal persistence. In the last group, two fresh readers still produce one acknowledgement per execution. This extends the recovery observation to a real database transaction path.

\begin{figure*}[t]
\centering
\includegraphics[width=0.98\textwidth]{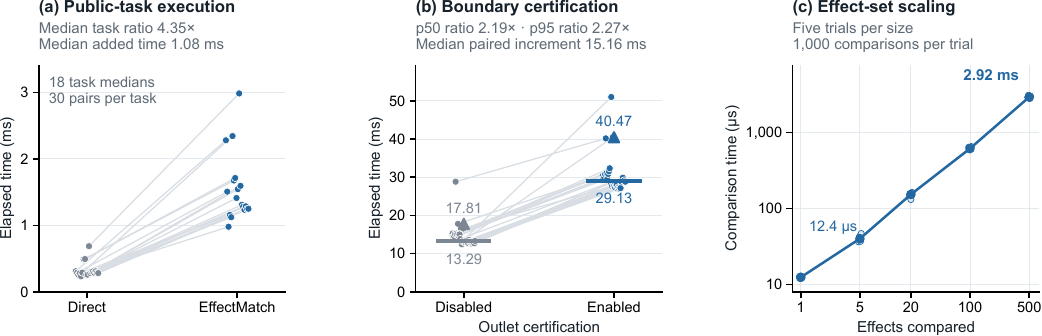}
\caption{Runtime cost at three levels. (a) Each connected pair shows one STATE task's median over 30 alternating direct/held pairs; annotations summarize the 18 task-level ratios and differences. (b) Thirty pairs of five-effect PostgreSQL executions isolate added outlet certification, retaining effect comparison and one-use permits in both conditions. Dots are runs, bars are medians, and triangles are nearest-rank p95. (c) In-memory comparison over 1--500 effects: small markers show five trials of 1,000 comparisons each, and the line joins trial medians.}
\label{fig:cost-scaling}
\end{figure*}

The original 80 cases test how these decisions apply to linear, fan-out, fan-in, and mixed dependency structures, crossing four disruptions with five instances per topology. A preflight blocker stops the plan before any effect; a mid-run failure preserves completed predecessors while stopping the failed node and dependent successors. In the mid-run failures, fan-in cases retain two completed predecessor effects per instance, while the other topologies retain one. Changed plans and stale or replayed approvals are rejected, and all 80 handoffs match durable state. Failure position and dependencies determine which work survives; the saved record carries that decision to subsequent execution.

\begin{table}[t]
\centering
\caption{Cross-backend outcomes. Counts are per variant; each operation or scenario is repeated as indicated.}
\label{tab:backend-outcomes}
\small
\setlength{\tabcolsep}{3pt}
\renewcommand{\arraystretch}{1.08}
\begin{tabularx}{\columnwidth}{@{}>{\raggedright\arraybackslash}p{.33\columnwidth}>{\raggedright\arraybackslash}p{.19\columnwidth}>{\raggedright\arraybackslash}X@{}}
\toprule
\textbf{Backend / coverage} & \textbf{Clean} & \textbf{Perturbed} \\
\midrule
MySQL audit prototype\newline 8 operations $\times$ 5 & 40/40 correct commits & Extra write: 40/40 rejected\newline Omission: 40/40 rejected \\
\addlinespace[4pt]
ToolSandbox held runtime\newline 4 scenarios $\times$ 3 & 12/12 correct commits & Extra reminder: 12/12 rejected\newline Omission: 12/12 rejected \\
\addlinespace[4pt]
GitHub readback\newline 8 operations $\times$ 5 & 40/40 fields match & Changed title: 40/40 mismatches detected; blocked\newline No readback: 40/40 blocked \\
\bottomrule
\end{tabularx}
\par\smallskip
\begin{minipage}{\columnwidth}
\footnotesize
MySQL and GitHub use predeclared operation cases; ToolSandbox uses public tasks with request-derived approval. MySQL implements comparison in a database-side audit procedure. GitHub reports registered-field readback; blocking means withholding the local continuation acknowledgement.
\end{minipage}
\end{table}

\subsection{RQ5: Cross-Backend Behavior, Deployment Conditions, and Cost}
\label{sec:rq5}

The preceding questions examine outcome checking, component contributions, live-agent integration, and continuation after failures. RQ5 asks how these checks transfer across backends, how backend control affects the response to a mismatch, and what the checks cost. Table~\ref{tab:backend-outcomes} extends pre-persistence checking to MySQL and ToolSandbox and adds a remote-state readback study on GitHub; Figure~\ref{fig:cost-scaling} separates public-task execution cost, PostgreSQL boundary-certification overhead, and effect-set comparison.

\emph{Checking before persistence.} MySQL's audit prototype and ToolSandbox's held runtime compare candidate results before saving them; rejecting a mismatch preserves the initial state. Direct MySQL transactions commit every extra-write and omitted-statement variant: atomic execution alone does not decide whether the resulting changes were approved. In ToolSandbox, the extra reminder still passes all 12 native task-goal checks under direct execution. Comparing the approved and observed state changes also detects this unrequested action, even when the requested goal is satisfied.

\emph{Remote-state readback and continuation.} GitHub provides an auxiliary test of remote-state readback and local continuation control. A fresh GET after the modifying request retrieves the registered issue fields: title, body, state, labels, and assignees. Matching fields produce a local continuation acknowledgement; an out-of-band title change or unavailable readback produces a saved decision that blocks this acknowledgement and automatic resend. These measurements concern the state at readback, without attributing the title change to the evaluated request. Previously applied remote changes remain in place.

\emph{Certifying execution conditions.} The component study checks admission and revalidation across 21 constructions, each repeated five times, covering foreign-database, host/network, asynchronous-signal, configuration-drift, and continuation conditions. The integrated PostgreSQL~17.6 study then checks five scenarios with 20 trials each; independent database readback agrees with runtime decisions in all 100 trials. Clean operations commit, effect comparison rejects additional writes, initial certification refuses an exposed host capability, fresh final-gate metadata catches a later capability grant, and a consumed permit prevents a second commit. Role grants expose the tested host capability for metadata inspection without invoking a host-file operation. These cases separate checking the business outcome from checking the execution conditions under which that outcome is observed.

\emph{Public-task execution cost.} Figure~\ref{fig:cost-scaling}(a) measures 30 alternating direct/held pairs for each of 18 STATE tasks, totaling 1,080 timed executions, all with correct results. Across task medians, the median held/direct ratio is 4.35$\times$ and the median added time is 1.08\,ms. Direct task medians are below one millisecond, so a small absolute increment still produces a substantial ratio. Timing includes native replay, runtime construction, comparison, permit handling, and durable SQLite state selection, with input preparation and two warmups per task and condition excluded. It measures the local execution path, excluding model calls and remote-service latency.

\emph{Boundary-certification cost.} Figure~\ref{fig:cost-scaling}(b) alternates outlet certification off and on for 30 pairs of the same five-effect PostgreSQL operation. Both paths retain relational catalog checks, effect comparison, and one-use permits; two warmup pairs are excluded. Timing covers preparation through commit, including metadata connections, with fixture setup and independent readback outside the timer. Certification raises median latency from 13.29 to 29.13\,ms (2.19$\times$), with a median per-pair increment of 15.16\,ms. Initial and final certificate capture take 7.01 and 7.21\,ms at the median, respectively, each opening a fresh metadata connection. Their contribution identifies connection and metadata access as the main optimization target while retaining a fresh final check.

Figure~\ref{fig:cost-scaling}(c) shows approximately linear growth in comparison time over the tested range of 1--500 effects, reaching a median of 2.92\,ms at 500 effects. This measurement isolates the computational cost of effect comparison; the execution paths also include staging, persistence, and metadata access.

\subsection{Scope and Deployment Implications}
\label{sec:scope}

\emph{Exactness within a certified profile.} Execution-level exactness requires the profile's declared coverage and execution attribution at its chosen comparison granularity (Section~\ref{sec:objects}). The GitHub study separately evaluates registered-state readback and local continuation control. The tested PostgreSQL execution path additionally requires its outlet certificate to remain valid through the final gate. The certification matrix covers five representative PostgreSQL boundary classes, not an exhaustive enumeration of database, extension, or operating-system behavior. For opaque backends, an already durable effect can be classified and contained when the required evidence is available; undoing it requires separately authorized compensation.

\emph{Authority and extension.} Deployment requires application-supplied approval and the trusted components in Section~\ref{sec:prototype}. Adapters provide their profiles' declared observation interfaces and execution controls, enabling outcome comparison and commit or continuation decisions. Gateway-mediated services also require control over credential and egress paths to route protected requests through the gateway. Coverage expands by specifying approved effects for new operations, registering observers, and validating adapter capabilities under the same contract. Cost results describe the tested execution paths.

\section{Related Work}
\label{sec:related}

Agent execution safety spans three related questions: what operation may be attempted, what effects that operation establishes, and whether later work may proceed. Prior work approaches this chain through threat benchmarks, runtime policy, authority lifecycle, and transactional recovery. Table~\ref{tab:positioning} groups representative systems by the object their primary contract controls and follows that object to the relevant execution boundary and outcome.

\begin{table}[t]
\centering
\caption{Execution contracts by primary control object.}
\label{tab:positioning}
\scriptsize
\begin{tabular}{@{}p{1.55cm}p{2.25cm}p{3.75cm}@{}}
\toprule
\textbf{Contract} & \textbf{Controlled object} & \textbf{Decision and retained result} \\
\midrule
Action/request authority~\cite{qin2026airguard,ma2026secureclaw} & proposed action or canonical request & authorize at tool admission or trusted-sink dispatch \\
Information flow~\cite{debenedetti2025camel,costa2025fides} & capability-controlled or labeled flow & permit or block the flow reaching tool use \\
Runtime rules~\cite{wang2026agentspec} & trigger events, predicates, and interventions & check and intervene at specified execution events \\
Authority lifecycle~\cite{santosgrueiro2026temporary,xu2026caplease} & authority witness or authorization instance & validate at commit; CapLease retains durable issue/prepare/commit state \\
Agent transactions~\cite{chen2026cordon,mohammadi2026atomix,chang2025sagallm} & staged effects and execution footprint & validate, commit, roll back, compensate, or recover \\
\sys{} & approved and observed outcomes of one execution & persist the comparison decision to govern commit, retry, and dependent work \\
\bottomrule
\end{tabular}
\end{table}

\emph{Agent risks and state-based evaluation.} AgentDojo evaluates prompt injection against agents operating tools over untrusted data, while ToolEmu uses an LM-emulated sandbox to expose long-tail execution risks~\cite{debenedetti2024agentdojo,ruan2024toolemu}. Agent Security Bench broadens the attack surface across prompts, tools, and memory, and AgentHarm tests whether an agent can carry a harmful request through a coherent multistep trajectory~\cite{zhang2025asb,andriushchenko2025agentharm}. Beyond attack labels, $\tau$-bench scores policy-guided tool use by comparing the final database state with an annotated goal state and measures reliability over repeated trials~\cite{yao2025taubench}. These benchmarks expose execution risks and establish the value of state-based assessment. \sys{} places outcome comparison in the protected execution path so that its result directly governs commit and subsequent work.

\emph{Policy and runtime guardrails.} CaMeL and Fides control how untrusted data may influence tool use through capability separation or information-flow constraints~\cite{debenedetti2025camel,costa2025fides}. AgentSpec expresses runtime constraints through structured rules comprising trigger events, predicates, and interventions; Progent enforces programmable least privilege over tool calls; ShieldAgent verifies action trajectories against explicit safety policies; and DRIFT combines dynamic rule checking with injection isolation~\cite{wang2026agentspec,shi2025progent,chen2025shieldagent,li2025drift}. These methods provide different mechanisms for expressing and enforcing execution policy. \sys{} studies a specific outcome constraint: compare actual changes with the result approved for this execution, within a boundary whose observation conditions have been established, and persist the decision governing commit and continuation.

\emph{Authority, consent, and proof.} AIRGuard derives step-level authority and intercepts normalized actions, while SecureClaw authorizes a canonical request at a trusted sink~\cite{qin2026airguard,ma2026secureclaw}. Temporary Authority revalidates freshness, dependency, and target evidence at commit; CapLease makes authorization issue, preparation, and consumption durable across retry and replay~\cite{santosgrueiro2026temporary,xu2026caplease}. Consent Integrity binds an approved presentation to the action that reaches a black-box agent boundary, and proof-oriented proposals carry planning or action certificates into later execution and audit~\cite{weng2026consent,cohen2026witness,wang2026pcaa}. UCON models ongoing conditions and mutable attributes during use, while macaroons attenuate delegated authority with contextual caveats~\cite{park2004ucon,birgisson2014macaroons}. These approaches refine the source, object, and validity conditions of authority. \sys{} binds the application-approved outcome to the current state and one execution, and checks that the approval still applies when comparing outcomes.

\emph{Transactions, effects, and recovery.} Cordon stages task effects in shadow state and an effect outbox, then validates the composed execution flow before commit; Atomix records read/effect footprints and delays settlement until earlier conflicting work can no longer arrive; SagaLLM combines validation logs, checkpoints, and compensation for multi-agent workflows~\cite{chen2026cordon,mohammadi2026atomix,chang2025sagallm}. Verification-aware tool wrappers use postcondition checks, idempotency keys, and verify-before-retry to reduce duplicate remote actions under non-atomic failures~\cite{mansoor2026verified}. Durable Functions gives stateful serverless workflows replay-based semantics, while AFT interposes on storage to provide atomic visibility across retried functions~\cite{burckhardt2021durable,sreekanti2020aft}. These systems extend foundations in effects, transactions, optimistic validation, sagas, and idempotency~\cite{lucassen1988effects,gray1992transaction,mohan1992aries,kung1981occ,garcia1987sagas,helland2007life,helland2012idempotence}. \sys{} takes the application's prior approved outcome as the comparison target and connects outcome matching, current approval, and the continuation decision to the same execution.

\emph{Persistent-outcome validation.} The protocol combines transactional staging, outcome comparison, and persistent decision records to link application approval, the checked candidate, and continuation to the same execution. Backends supply either a staging-and-commit control point or request dispatch followed by readback; the protocol maintains the correspondence among approval, outcome, and continuation at these points. It builds on provenance~\cite{buneman2001why,green2007provenance,cheney2009provenance}, capability systems~\cite{dennis1966capabilities,saltzer1975protection}, runtime enforcement~\cite{schneider2000enforceable,ligatti2005edit}, and sender-constrained credentials~\cite{rfc9449dpop}.

\section{Conclusion}
\label{sec:conclusion}

The persistent changes left by an agent operation must conform to the approval for that execution. Within a controlled execution boundary, \sys{} binds application approval to the current state and one execution, checks actual changes, and persists the decision governing commit and subsequent work. The Effect Commit Contract connects approval, outcome comparison, and continuation so that the execution decision remains available after process exit or recovery.

On twelve public business tasks with approved outcomes independently specified before execution, \sys{} preserved all clean executions and rejected the mismatches in all 39 task--fault combinations, each repeated three times. Extended evaluation covered reference-based comparison on 206 public tasks, live execution with three agent configurations, cross-process recovery, and different database and external-service backends. By making continuation depend on the persisted outcome decision, the protocol prevents executions that violate approval or remain unresolved from propagating as success.

\section*{Acknowledgment}
OpenAI Codex and Anthropic Claude were used as author-directed tools for writing and revision support, code development, and schematic preparation. The authors reviewed all outputs and take full responsibility for the manuscript.

\bibliographystyle{ieeetr}
\bibliography{verdict_tse_references}

\end{document}


\title{Supplementary Material for\\Beyond Approved Actions: Runtime Validation of Persistent Outcomes in Agent Workflows}
\author{Haoran~Zhang, Hengtong~Zhang, Zhiyu~Liang, Yu~Yan, Decheng~Zuo,
  and~Hongzhi~Wang,~\IEEEmembership{Senior Member, IEEE}%
\thanks{The authors are with the Faculty of Computing, Harbin Institute of Technology, Harbin 150001, China. E-mail: zhr@stu.hit.edu.cn, \{hengtong, zyliang, yuyan, zuodc, wangzh\}@hit.edu.cn. Corresponding author: Hongzhi Wang.}}
\markboth{Supplementary Material}{Runtime Validation of Persistent Outcomes in Agent Workflows}
\maketitle

This supplement provides construction examples, implementation details, full comparison results, and live-agent settings for \sys{}. Sections~\ref{sec:ae} and~\ref{sec:boundary} accompany the execution model and protocol in Sections~II--III of the main paper. Sections~\ref{sec:experiments} and~\ref{sec:live} accompany its evaluation. All results come from the experiments reported in the paper; tables here expose additional cells or task-level detail. Section, table, and equation numbers beginning with S refer to this document.

\section{Constructing Approved and Observed Outcomes}
\label{sec:ae}

\subsection{Inputs, Representation, and Independence}

An approved outcome $A$ fixes what the application accepts before the candidate executes. An observed outcome $E$ represents the changes established from execution evidence. Both use the granularity of the registered profile $p$.

For a write-event profile, entries retain operations, targets, and multiplicities. For a net-state profile, entries describe changes between the initial and candidate states, retaining values or value digests. Thus two writes that restore the initial value can differ under event comparison while producing an empty net-state difference. The profile, rather than the trial outcome, selects this interpretation.

The independently approved experiments construct their targets from public task requests, initial business data, and hand-authored task specifications with independent business calculations. PostgreSQL contracts calculate requested row values without executing the candidate SQL. STATE contracts apply separate Python transformations without invoking native mutating handlers. The runtime receives the resulting approved content and its execution binding. Benchmark target constraints and post-execution scores are reserved for assessment after the durable decision. The 206-task STATE comparison has a different target source. It saves one separate clean native replay per task before running the compared conditions. That saved result supplies the reference for outcome checking. The comparison measures enforcement with a supplied target; the independently specified subset measures enforcement with a request-derived target. These populations retain their respective approval sources throughout the evaluation.

For the snapshot adapters, the approved target is a complete expected business snapshot. Its difference from the initial snapshot gives the approved effect representation. The candidate snapshot is independently produced by native tools, and its difference from the same initial snapshot gives the observed representation. Unrequested business records retain their initial values in the target, so changing an unrelated field creates a mismatch. The comparison operates over the registered business namespaces; conversation transcripts are not treated as business-state effects.

\subsection{Example: Partial Order Cancellation}

The public customer-support task \id{150-spare_cancel_in_transit_fee_decline} requests cancellation of the still-cancellable part of order \id{ORD-7281}. Its initial state contains a delivered phone and an unshipped phone case. The independent specification selects the unshipped case, preserves the delivered phone, and derives the cancellation and refund fields from the initial order. Table~\ref{tab:approval-example} shows the resulting changes. Amounts use the task's currency units.

\begin{table}[t]
\centering
\caption{Request-derived changes for order ORD-7281. All other registered business fields remain unchanged.}
\label{tab:approval-example}
\small\setlength{\tabcolsep}{3pt}
\begin{tabularx}{\columnwidth}{@{}Yll@{}}
\toprule
\textbf{Object / field} & \textbf{Initial} & \textbf{Approved}\\
\midrule
Order: status & partially delivered & partially cancelled\\
Case item: status & confirmed & cancelled\\
Case item: refund amount & null & 35\\
Case item: refund method & null & split\\
\bottomrule
\end{tabularx}
\end{table}

The order item is \id{ITEM-10286}. The calculation sets its refund to its initial unit price, 35, and its refund method to the order's existing split-payment method. The order becomes partially cancelled because the delivered item remains unchanged. A native execution that produces these changes matches the target.

In the controlled altered-value variant, the harness increases the refund amount in the candidate snapshot by one after native replay. The resulting candidate still contains the requested cancellation, but its refund is 36. Value-sensitive comparison rejects the candidate and preserves the initial durable state. The injected value is an experimental fault; the runtime receives candidate state, not the fault label. This example also shows why the requested operation alone does not define $A$. A generic ``cancel order'' call does not specify which already-delivered item must remain, the permitted refund amount, or the required payment method. The independent application specification supplies these constraints before execution.

\subsection{Example: Database Migration and Hidden Writes}

For MCPMark's \id{customer_data_migration}, the approval calculation reads the public import records and the initial database. It constructs the 200 requested customer rows, including their field values and defaults, before candidate SQL executes. Unrequested registered rows must remain unchanged. The evaluated command runs in a private PostgreSQL transaction, held open for inspection. The observer reads the registered relations before execution and in the held candidate, then reconciles those differences with protected, transaction-local write records. A hidden fixture trigger can additionally update the \id{Genre} relation while the submitted SQL stays unchanged. Because \id{Genre} belongs to the registered scope, its change enters $E$ despite being absent from $A$. Missing-customer and altered-name fixtures instead remove or change an approved result. All three variants reach the same outcome-comparison step; rejection rolls back the held transaction.

The invoice illustration in Figure~2 of the main paper uses a smaller relation-write projection: one invoice update is approved, and a notification insertion is additional. The public database contracts retain row values as well. Relation-operation counts alone cannot distinguish two 200-row imports that differ only in a customer's name.

\subsection{Execution Binding and Stored Decisions}

The approval binding $B_A$ associates approved content with its source, task state, plan revision, profile, and execution occurrence. The final permit $B_G$ associates a successful check with the particular held candidate or mediated request that may be committed or dispatched. A signed permit for one candidate cannot be transferred to another candidate merely because both belong to the same task. The occurrence and permit stores retain one-use consumption across subsequent attempts.

The outcome classification and stored disposition describe different facts. \id{EXACT}, \id{DIVERGED}, and \id{INDETERMINATE} classify outcome evidence under the profile. The journal records whether execution committed, was rejected before commitment, or requires resolution after possible execution. For example, the invoice candidate is divergent, but its saved disposition is rejection before commit.

\section{Observation Boundary and Backend Realization}
\label{sec:boundary}

\subsection{What the PostgreSQL Deployment Registers}

The PostgreSQL adapter operates over an explicitly provisioned configuration. Registration specifies the business relations, supported operation entry points or DML, trigger and function definitions, execution role, audit objects, and deployment restrictions.

The public three-task path snapshots 11--13 registered relations, depending on the task. This includes relations that a task does not request to change. It does not discover its observation scope from the rows mentioned in the current approval. A trusted metadata connection reads the live catalog. The execution connection retains restricted privileges; administrative metadata access does not become an agent capability.

Boundary checks classify supported routes as observed or denied and refuse an unclassified route. Relevant inputs include relation and trigger definitions, callable functions and their permissions, role membership, database TEMP privileges, extension and foreign-data exposure, and the configured host/network restrictions. Application functions are admitted only within the supported registered configuration; this is not an unrestricted analysis of arbitrary function code.

Write-recording triggers attach stage identity to changes in the registered relations. The adapter forces deferred constraint triggers using \texttt{SET CONSTRAINTS ALL IMMEDIATE} before collecting evidence. It locks the registered relations before reading the initial state and keeps those locks until the held transaction commits or rolls back. Candidate-state and audit checks then concern the same transaction selected for promotion. Administrative grants, function implementations, and external isolation are deployment-managed configuration and must remain stable during protected execution. The final gate obtains fresh metadata and rejects visible changes relative to the certified configuration. This recheck supplements the relation locks: the locks stabilize compared data, while fresh certification detects configuration changes visible at the gate.

Snapshot cost over the 11--13 registered relations is part of this prototype's observation work; the separate in-memory comparison benchmark measures only comparison computation.

\subsection{Boundary Test Inventory}

The component matrix contains 21 constructions, each repeated five times. Table~\ref{tab:boundary-components} reports the tested configuration classes. Its counts are satisfied expected-pass and expected-refusal assertions from admission, certificate, and continuation components. Represented host, network, foreign-database, and notification capabilities are examined without invoking those outlets. Harness-initialized safety counters are not used as measurements of independently monitored outlet activity.

\begin{table}[t]
\centering
\caption{Component-level boundary inventory. Each construction has five repetitions; counts report satisfied case assertions.}
\label{tab:boundary-components}
\small\setlength{\tabcolsep}{3pt}
\begin{tabularx}{\columnwidth}{@{}Yrrr@{}}
\toprule
\textbf{Configuration class} & \textbf{Cases} & \textbf{Pass} & \textbf{Refuse}\\
\midrule
Base / foreign database & 4 & 10 & 10\\
Host / network & 4 & 10 & 10\\
Asynchronous signals & 5 & 10 & 15\\
Function / configuration drift & 4 & 5 & 15\\
Continuation authority & 4 & 5 & 15\\
\midrule
Total & 21 & 40 & 65\\
\bottomrule
\end{tabularx}
\end{table}

The base/foreign class tests a stable certificate, a denied foreign outlet, a foreign table, and a callable foreign-database path. Host/network cases pair denied capabilities with an exposed host role or unsupported host-facing extension. Signal cases include a benign statement, a registered benign function, and direct or wrapped notification paths. Function cases change public-execute permission, search path, or execute ACL relative to stable configuration. Continuation cases compare a registered continuation with an independent reader, wrong authority lineage, and a replay.

A separate integrated study connects certification to actual held transactions and independent durable-state readback. It uses five scenarios with 20 trials each: a clean five-effect operation, an extra sibling outbox write, an exposed host capability before preparation, a host-capability grant after preparation, and reuse of a consumed permit. All 100 trials match their runtime and durable-state expectations. Clean operations commit; the extra-write and certification failures preserve initial application rows; replay leaves only the first approved result. The replay scenario contains two execute attempts per trial. Host-capability grants are inspected as catalog changes, without executing a host-file primitive.

\subsection{What Each Backend Implementation Establishes}

The implementations expose different control points:
\begin{itemize}
\setlength{\itemsep}{2pt}
\item \emph{PostgreSQL:} private native transaction, registered relation snapshots and protected audit, boundary certification, final rechecking, and one-use promotion.
\item \emph{STATE and ToolSandbox:} native business tools produce a private simulator snapshot. The held runtime checks that snapshot before selecting durable state in SQLite. These adapters cover simulator business state.
\item \emph{MySQL:} a database-side audit procedure compares predeclared row transitions before committing a native transaction. This is the evaluated MySQL adapter prototype.
\item \emph{GitHub:} a fresh GET checks registered issue fields after the modifying request. A saved local decision controls the continuation acknowledgement. It neither rolls back remote changes nor establishes attribution of all effects to that request.
\end{itemize}

The GitHub fields are title, body, state, labels, and assignees. A matching readback, an out-of-band title change, and unavailable readback are distinct study conditions. Historical classifier strings containing ``exact'' or ``diverged'' in these records denote field-match categories for this runner. The reported evidence is registered-field readback and local continuation control, as in the main paper's RQ5.

\section{Experimental Populations, Comparisons, and Additional Results}
\label{sec:experiments}

\subsection{Public Tasks and Selection}

MCPMark supplies the database tasks, STATE-Bench supplies shopping, customer-support, and travel tasks, and ToolSandbox supplies four single-action scenarios. Their public repositories are listed below.\footnote{MCPMark: \url{https://github.com/eval-sys/mcpmark}; STATE-Bench: \url{https://github.com/microsoft/STATE-Bench}; ToolSandbox: \url{https://github.com/apple/ToolSandbox}.}

The twelve independently approved STATE tasks in Table~\ref{tab:state-tasks} are a purposeful, request-resolvable subset with four tasks per domain. Requests with unresolved choices or missing subsequent consent are not used to construct these specifications. Inputs include the released user messages preceding the first mutating invocation and the initial business data. Native \id{confirm=false} previews are not counted as mutations in this independent subset. Approval generation is hand-authored application integration, not an evaluated language-model extraction task.

\begin{table*}[t]
\centering
\caption{The twelve independently specified STATE tasks. An asterisk identifies the six live-agent tasks. Task IDs are local to the named domain.}
\label{tab:state-tasks}
\small\setlength{\tabcolsep}{4pt}
\begin{tabularx}{\textwidth}{@{}p{1.65cm}p{9.2cm}Y@{}}
\toprule
\textbf{Domain} & \textbf{Released task ID} & \textbf{Approved operation}\\
\midrule
Shopping & \id{10-loyalty_points_on_discount} & Add laptop with SAVE10 discount\\
& \id{14-price_drop_disclosure} & Add the specified laptop\\
& \id{20-triple_policy_compound_disclosure}$^{*}$ & Add two products; gift-wrap only the speaker\\
& \id{84-remove_qty_scope_all_vs_one}$^{*}$ & Remove the entire webcam cart line\\
\addlinespace
Support & \id{11-cancel_before_shipment}$^{*}$ & Cancel an unshipped order\\
& \id{14-cancel_split_payment} & Cancel an order with split payment\\
& \id{150-spare_cancel_in_transit_fee_decline}$^{*}$ & Cancel only the unshipped item\\
& \id{17-cancel_price_match} & Refund the price difference\\
\addlinespace
Travel & \id{2-cancel_business_international}$^{*}$ & Cancel two flights and the associated hotel\\
& \id{100-challenge_medical_yyze_intl_lt7d} & Change flight using the medical-fee rule\\
& \id{103-change_business_international_fare_diff_only} & Change international business-class flight\\
& \id{104-change_domestic_gt7d_personal_fee_tier}$^{*}$ & Change domestic flight; preserve preferences\\
\bottomrule
\end{tabularx}
\end{table*}

The broader STATE comparison retains all 206 historically eligible write-task candidates from the frozen inventory. A new native inventory pass covered 300 tasks with three replays each; it did not redefine the 206-task cohort using the new comparative outcomes. Nine tasks in that cohort perform writes but finish with no net state change. Their empty change targets remain in the population. Omitting or duplicating a call can also leave the final result unchanged, so injected variants are retained whether or not they cause a mismatch.

The three independently approved PostgreSQL tasks are \id{customer_data_migration}, \id{employee_hierarchy_management}, and \id{sales_and_music_charts}. Their specifications calculate customer rows, employee/report/customer assignments and salaries, and monthly invoice aggregates and music rankings, respectively.

Schema initialization is separately approved setup outside the held DML transaction. For the employee task, the detailed add-two/delete-one instructions and public verification query imply nine employees, while a final sentence says ten; the experiment explicitly adopts the detailed instructions. Ranking ties use source-name collation order read before execution, with quantities and revenues independently calculated.

The remaining eighteen tasks in the 21-task, 269-SQL-call MCPMark inventory have explicit implementation dispositions: ten need setup, namespace, sequencing, or business-contract support; five request functions, triggers, or privilege policies as outputs; one concerns physical index design; one concerns permission-catalog auditing; and one needs a vector-extension profile. These are unimplemented task/profile combinations, not measured successful executions or runtime rejections.

\subsection{Conditions, Fault Placement, and Denominators}

Direct PostgreSQL execution already uses a transaction and aborts on SQL errors. The AgentSpec condition uses the original \id{Rule}, \id{RuleState}, \id{RuleInterpreter}, \id{Action}, and \id{Stop} implementation from AgentSpec 0.1.0, with a custom deterministic SQL target predicate registered through its \id{predicate11} slot. The rule checks visible write operations and target relations against the application's relation-operation policy. It rejects the separate explicit-extra-write control in all nine runs. Hidden-write fixtures leave submitted SQL unchanged and modify the resulting state through backend triggers. The comparison concerns this call-level rule configuration.

For STATE, the error-reject baseline uses the same native handlers in a private copy and discards that copy if a tool raises, returns a nonempty error field, or reports \id{success=false}. The integrated condition compares the candidate outcome with approval before saving. Its effect-comparison ablation retains staging and permit handling but removes this comparison. Tool response errors and outcome mismatches are measured separately.

PostgreSQL fault fixtures add a \id{Genre} write, omit a requested row, or alter a requested value. STATE omission and duplication modify the first selected write invocation. For the independently approved STATE subset, the two additional faults are candidate-snapshot injections after native replay: add one to the relevant cart total, refund amount, or booking price; or change an account email. The adapter and policy do not receive the fault label.

The twelve-task experiment has 20 effect cells and three additional binding cells per task, each repeated three times, yielding 828 executions. PostgreSQL has 16 effect cells and six binding/boundary cells per task, also repeated three times, yielding 198 executions. The 206-task comparison has three variants, four conditions, and three repetitions per task: 7,416 executions. Table~\ref{tab:conditional-full} includes the ablated condition omitted from the main comparison table. Of 618 omission runs, 354 produce mismatching candidates; of 618 duplication runs, 138 do so. The remaining injected runs are still counted. Repetitions use fresh execution state, but repeated deterministic calls are not additional task identities or independent samples of agent behavior.

\begin{table}[t]
\centering
\caption{Full 206-task comparison. Each column has 618 runs per condition. Clean counts matching durable states; fault columns count mismatching commits.}
\label{tab:conditional-full}
\small\setlength{\tabcolsep}{3pt}
\begin{tabularx}{\columnwidth}{@{}Yrrr@{}}
\toprule
\textbf{Condition} & \textbf{Clean $\uparrow$} & \textbf{Omit $\downarrow$} & \textbf{Duplicate $\downarrow$}\\
\midrule
Direct & 618 & 354 & 138\\
Error-reject & 597 & 138 & 138\\
\sys{} & 618 & 0 & 0\\
No result comparison & 618 & 354 & 138\\
\bottomrule
\end{tabularx}
\par\smallskip
\begin{minipage}{\columnwidth}\footnotesize
Matching durable state is distinct from issuing a commit. Error-reject saves 591 clean candidates; six other clean runs retain an initial state already equal to the reference, giving 597 matching states.
\end{minipage}
\end{table}

\subsection{Full Mechanism Comparison}

Table~\ref{tab:mechanism-full} expands the four scenarios shown in the main paper to all eight frozen scenarios. Each of the six conditions has 20 fresh runs per scenario, totaling 960 executions on local PostgreSQL, SQLite, and loopback HTTP fixtures. Direct and Allowlist are separate conditions. Robust Workflow adds constrained requests, current-state checks, readback, and transaction/retry controls. NativeAudit adds ordinary backend audit evidence. Complete Audit is an instrumented observation oracle with the full protected-effect set and a locally derived expected set. \sys{} additionally binds the approved source and execution context.

\begin{table*}[t]
\centering
\caption{All eight mechanism scenarios. Every displayed outcome occurred in 20/20 runs in its cell. Direct and Allowlist were executed separately. Complete Audit is the instrumented observation oracle.}
\label{tab:mechanism-full}
\small\setlength{\tabcolsep}{4pt}\renewcommand{\arraystretch}{1.12}
\begin{tabularx}{\textwidth}{@{}p{3.2cm}YYYYYY@{}}
\toprule
\textbf{Scenario} & \textbf{Direct} & \textbf{Allowlist} & \textbf{Robust Workflow} & \textbf{NativeAudit} & \textbf{Complete Audit} & \textbf{\sys{}}\\
\midrule
Clean chain & Exact & Exact & Exact & Exact & Exact & Exact\\
Extra PostgreSQL write & Extra commit & Extra commit & Extra commit & Reject & Reject & Reject\\
Mediated hidden field & Extra commit & Extra commit & Exact & Exact & Exact & Exact\\
Stale selector & Stale commit & Stale commit & Reject & Reject & Reject & Reject\\
Opaque remote extra & Extra; advance & Extra; advance & Extra; advance & Uncertain; stop & Diverged; stop & Diverged; stop\\
Readback unavailable & Response; advance & Response; advance & Reconcile & Uncertain; stop & Uncertain; stop & Uncertain; stop\\
Response loss / restart & Resend; duplicate & Resend; duplicate & Reconcile & Reconcile & Reconcile & Uncertain; stop\\
Wrong-source target & Accept local match & Accept local match & Accept local match & Accept local match & Accept local match & Reject\\
\bottomrule
\end{tabularx}
\par\smallskip
\begin{minipage}{\textwidth}\footnotesize
Exact: approved result is realized. Reject: no candidate promotion. Reconcile and uncertain: no dependent release or automatic resend. ``Extra; advance'' leaves one unapproved remote effect and runs one dependent action; ``diverged; stop'' and ``uncertain; stop'' in that row leave the same remote effect but block continuation. Direct/Allowlist resend once in the response-loss row, causing one extra effect. Wrong-source matching preserves the locally expected effect but lacks the designated approval source. These labels summarize recorded behavior; they do not introduce new protocol states.
\end{minipage}
\end{table*}

The oracle's observation advantage explains its effect-level results. In the wrong-source scenario, the effects still match the locally supplied target, so observing them more completely cannot establish that the target came from the designated approver. Conversely, once a remote effect has occurred, a later blocked continuation does not undo it. The matrix keeps these outcomes separate from pre-commit rejection.

\subsection{Ablation Interventions}

The six dedicated fault pairs in Table~\ref{tab:ablation-details} each contain 20 full and 20 ablated executions. Candidate binding and fresh-certificate capture also have 20 clean controls per condition. These are 240 fault executions plus 80 clean controls. Public-task comparison and binding ablations use their own task populations and three repetitions, as specified above.

\begin{table}[t]
\centering
\caption{Dedicated ablations: removed mechanism and the failure observed in all 20 ablated fault runs. Full execution rejects or blocks all 20 paired faults.}
\label{tab:ablation-details}
\small\setlength{\tabcolsep}{3pt}
\begin{tabularx}{\columnwidth}{@{}p{2.4cm}Y@{}}
\toprule
\textbf{Removal} & \textbf{Discriminating intervention / outcome}\\
\midrule
Complete observation & Omit the additional write from $E$; extra effect commits.\\
Approval-source check & Substitute the source while effects match; wrong-source approval accepted.\\
One-use execution binding & Reuse authority; an additional external effect occurs.\\
Durable hazard record & Lose saved divergence; a dependent action executes.\\
Candidate binding & Present a paid-candidate permit for a cancelled candidate; replacement commits.\\
Fresh boundary capture & Reuse the initial certificate after a TEMP grant; stale-boundary execution commits.\\
\bottomrule
\end{tabularx}
\end{table}

Candidate-binding removal disables the runtime's stage-identity checks and the adapter's permit-to-candidate tuple check. Signature, expiry, current policy, base-state validation, candidate-to-own-observation validation, and one-use consumption remain active. The fixture first validates a file-backed invoice candidate with status paid, then presents its permit for a different candidate with status cancelled. Independent file readback identifies the committed candidate.

Fresh-boundary removal returns the genuine initial certificate at the final gate instead of recapturing metadata. The fault grants database TEMP permission after staging. The retained relational catalog check does not inventory that database-level privilege; full outlet recapture does. Both the catalog and business-effect checks remain active. The ablated result therefore still matches the approved five-effect deletion, but commits under a changed certification condition. No temporary table or host operation is invoked.

An earlier role-grant diagnostic was detected by both the outlet and catalog checks; it is retained separately and is not pooled into this discriminating TEMP-grant study. The public-task occurrence ablations remove occurrence/action and pre-state binding together. Their tested intervention changes occurrence while leaving the candidate business state correct. The public PostgreSQL boundary ablation removes the combined final boundary/audit recheck; the dedicated TEMP study isolates fresh certificate capture. These interventions have different scopes and are reported separately in the main paper.

\subsection{Recovery, Task-Level Timing, and Other Backends}

Recovery and timing use the same eighteen STATE tasks: the six lexicographically first eligible multistep tasks per domain. Table~\ref{tab:task-cost} identifies every task and gives the execution medians underlying the main cost figure.

Recovery launches one producer and two fresh consumers per case, with six normal/interrupted scenarios and five repetitions. Producers exit at the selected checkpoint; consumers read saved state and terminal records without reconstructing missing terminals. A unique harness acknowledgement represents release of dependent work. Thus the correct-state/no-terminal cases measure withheld continuation pending reconciliation.

\begin{table*}[t]
\centering
\caption{Task-level medians for the eighteen-task timing and recovery cohort. Each task has 30 alternating direct/held timing pairs. Times are milliseconds; the ratio divides the two task medians.}
\label{tab:task-cost}
\small\setlength{\tabcolsep}{4pt}
\begin{tabularx}{\textwidth}{@{}p{1.45cm}Yrrr@{}}
\toprule
\textbf{Domain} & \textbf{Released task ID} & \textbf{Direct} & \textbf{Held} & \textbf{Ratio}\\
\midrule
Support & \id{10-return_full_order} & .309 & 2.281 & 7.37\\
& \id{100-challenge_warranty_maxed_return_option} & .287 & 1.507 & 5.25\\
& \id{105-hard_price_match_after_exchange_request} & .256 & .979 & 3.82\\
& \id{107-hard_bundle_price_match_requires_approval} & .278 & 1.157 & 4.16\\
& \id{11-cancel_before_shipment} & .234 & 1.122 & 4.80\\
& \id{111-hard_exchange_downgrade_cash_demand} & .289 & 2.344 & 8.11\\
\addlinespace
Shopping & \id{10-loyalty_points_on_discount} & .489 & 1.675 & 3.42\\
& \id{100-goodwill_exception_fabrication_decline} & .494 & 1.713 & 3.47\\
& \id{111-hard_shipping_gold_express_vs_nextday} & .257 & 1.412 & 5.50\\
& \id{118-hard_loyalty_redemption_cap_preview} & .259 & 1.545 & 5.96\\
& \id{12-gift_wrap_silent_fee} & .687 & 2.985 & 4.34\\
& \id{138-hard_remove_desk_keep_invalid_promo} & .282 & 1.595 & 5.66\\
\addlinespace
Travel & \id{100-challenge_medical_yyze_intl_lt7d} & .303 & 1.309 & 4.31\\
& \id{103-change_business_international_fare_diff_only} & .308 & 1.281 & 4.16\\
& \id{104-change_domestic_gt7d_personal_fee_tier} & .286 & 1.235 & 4.32\\
& \id{106-change_domestic_weather_zero_fee} & .310 & 1.286 & 4.16\\
& \id{109-change_international_gt7d_bereavement_discount} & .287 & 1.246 & 4.35\\
& \id{110-change_domestic_schedule_change_zero_fee} & .282 & 1.250 & 4.43\\
\bottomrule
\end{tabularx}
\end{table*}

STATE timing includes native replay, runtime construction, comparison, permit handling, and durable SQLite selection. Input loading, clean-reference generation, initial store setup, resets, scoring, and diagnostic replay are outside the timer; two warmups per task and condition are excluded. All 1,080 timed executions produce correct results. The median across task-level held/direct ratios is 4.345, while the median added time across task medians is 1.081\,ms. The two statistics answer different questions and are computed from unrounded data.

PostgreSQL certification timing uses 30 alternating pairs of one five-effect operation, with two warmup pairs excluded. Both conditions retain relational catalog checks, effect comparison, and one-use permits. Initial and final outlet capture each opens a fresh metadata connection. The disabled/enabled total medians are 13.292/29.127\,ms; nearest-rank p95 values are 17.810/40.472\,ms. The median paired increment is 15.157\,ms. Fixture setup, fault installation, independent readback, and cleanup are outside this timer.

The separate 1--500-effect comparison measurement uses five trials of 1,000 comparisons per size and excludes staging and persistence.

MySQL's eight authored operations are account activation, bulk transition, balance transfer, order archival, inventory upsert, zero-balance deletion, account archival projection, and joined bonus update. Each runs clean, extra-write, and omitted-statement variants in direct and audit-procedure conditions, with five repetitions: 240 executions.

ToolSandbox uses \id{cellular_off}, \id{wifi_off}, \id{remove_contact_with_id}, and \id{update_contact_with_id_and_phone_number}. Clean, extra-reminder, and omitted-action variants each have direct and held conditions with three repetitions: 72 executions. ToolSandbox approval uses the user request and initial state; milestone scoring runs after the decision.

GitHub's eight shapes cover creation, title revision, body revision, closing, reopening, title-plus-body revision, body-plus-closing, and reopening-plus-body revision. Each has five repetitions under clean, out-of-band title-change, and unavailable-readback conditions: 120 executions. The study saves intended fields before the write and obtains observed fields through a fresh GET. Only matching readback releases the local acknowledgement; the other conditions block it without resending. These are operation cases on one object type, rather than eight public benchmark tasks.

\subsection{Execution Environments}

The public-task experiments used an Apple M5 host. Native simulator clients used Python~3.12.13. The integrated boundary studies used PostgreSQL~17.6 on aarch64 with psycopg~3.3.4 and a Docker image based on \id{pgvector:0.8.0-pg17-bookworm}; the image digest is retained in the experiment manifest. MySQL used a disposable 8.4 instance. Database fixtures ran on local internal Docker networks.

The boundary experiments inspected privileged capabilities through controlled metadata changes rather than invoking host or external-network operations. Model inference is confined to the live-agent and separately reported surface-approver studies; deterministic replays, recovery, certification, and timing require no model calls or GPU computation.

\section{Live-Agent Inputs, Configuration, and Finalization}
\label{sec:live}

\subsection{Shared Task Interface}

The six starred tasks in Table~\ref{tab:state-tasks} provide two operations per domain. Each is run with three configurations, two execution conditions, and three fresh repetitions, producing 108 episodes. Support task \id{11-cancel_before_shipment} was also used in the disclosed pilot; pilot results are excluded. The formal schedule is shuffled with seed 20260921 and permits up to three isolated episodes concurrently. Direct and integrated episodes generate their own call sequences.

Each configuration receives the native domain system prompt, formatted with task time and user identity; released user messages before the original trajectory's first mutation; and native business-tool schemas. Two additional tools request the fixed experimental customer's confirmation and finalize the task. The shared wrapper is:

\begin{quote}\small
Carry out the following customer request using only the provided business tools. The following are customer messages already received, in order; earlier confirmations apply only to their stated scope. Use the native preview/confirmation semantics. When customer confirmation is needed after preview, call request\_customer\_confirmation to obtain the reply, then continue. Call finish\_task before giving your final response. Do not inspect files, run shell commands, use other tools, or consult external sources.
\end{quote}

The message bundle follows this wrapper. No prior assistant/tool-result trace, expected snapshot, hidden target requirements, or scorer output is included. Some released travel messages already specify approved amounts; those amounts remain part of the public input. The approval specification is compiled and saved before inference, separately from the model-facing packet. Confirmation replies are fixed per task before execution and are released when the required native preview and target arguments are present. For the partial cancellation example, the reply approves cancelling only the unshipped phone case, reversing its payment to the original methods, and preserving the delivered phone. Reply eligibility uses preview metadata and arguments, not the expected state or benchmark score. This supplies a repeatable task interaction while preserving the stated scope of the customer's request.

\subsection{Models, Harnesses, and Budgets}

The minimal DeepSeek condition uses native function calls through the API. DeepSeek+DSH uses DSH~0.1.5-rc.2 with its SDK harness and native tools. Both request \id{deepseek-flash}, disable thinking, and cap output at 2,048 tokens per request. Codex CLI~0.154.0 requests \id{gpt-5.6-sol} with high reasoning. It runs in an ephemeral task workspace with only the episode's business-tool server enabled; shell, web, personal skills, memories, and delegation are disabled. Model identifiers are the requested identifiers recorded at execution time.

Both DeepSeek configurations have a 24-model-iteration cap. All configurations have a 40-native-business-call and 180-second episode bound; Codex does not have a verified per-inference iteration cap. Harness-specific instructions and role/request formatting remain in place, so these are three agent configurations rather than an isolated comparison of model capability.

The direct path saves native state after each call. The integrated path keeps one private task candidate until finalization, when the held runtime checks the whole task result against independent approval before saving it.

\subsection{Business-State Equivalence}

The comparison rule is fixed before formal outcomes. Existing object identifiers, quantities, prices, preferences, unrelated state, and item multiplicities remain checked. Newly allocated shopping cart-item IDs may differ if a one-to-one mapping preserves their business fields and all associated references. Cart item-list order is ignored. This handles system-assigned surrogate IDs without accepting different products, quantities, or wrapping choices.

Strict raw-snapshot equality is recorded separately. One Codex/direct shopping episode differs only under this permitted representation rule; it is business-equivalent to the independent target.

\subsection{Original Cohort and Targeted Follow-Up}

Table~\ref{tab:live-results} distinguishes candidate correctness, durable correctness, and finalization in the original 108 episodes. All candidate business states match their independent targets, and all episodes finish without process errors or timeouts. The batch therefore measures successful-task integration; controlled fault studies supply the incorrect-result prevention evidence.

\begin{table}[t]
\centering
\caption{Original live-agent cohort. Each row contains 18 episodes; candidate and durable columns count matching business states.}
\label{tab:live-results}
\small\setlength{\tabcolsep}{3pt}
\begin{tabularx}{\columnwidth}{@{}Ylrrr@{}}
\toprule
\textbf{Configuration} & \textbf{Path} & \textbf{Candidate} & \textbf{Durable} & \textbf{Finalized}\\
\midrule
DeepSeek API & Direct & 18 & 18 & 15\\
& Integrated & 18 & 15 & 15\\
DeepSeek+DSH & Direct & 18 & 18 & 18\\
& Integrated & 18 & 18 & 18\\
Codex & Direct & 18 & 18 & 18\\
& Integrated & 18 & 18 & 18\\
\bottomrule
\end{tabularx}
\end{table}

The three unfinished integrated results are repetitions of the selective-gift-wrap shopping task. Their correct candidates remained private because the API agent omitted \id{finish_task}.

The host-finalization follow-up reruns exactly those three episodes with unchanged public packets, independent approvals, model settings, and budgets. On normal completed turns, the host invokes the existing checker if the model has not finalized; errors, timeouts, interruptions, and exhausted budgets do not trigger this hook. All three fresh candidates and durable states are correct. The main paper's marked API cell combines the 15 original successes with these three targeted successes; the original cohort remains recorded separately here.

The original run made 645 native business-tool calls and 77 confirmation requests. Provider-reported usage was 5,457,463 input tokens, including cached input, and 80,697 output tokens; the three follow-ups used 45,191 input and 1,492 output tokens. These counts describe model/harness usage, not the runtime overhead measured in Section~\ref{sec:experiments}.

Per-episode records preserve public inputs, precomputed approvals, model events, native calls, candidate and durable snapshots, and available finalization receipts.